\documentclass[10pt, conference, compsocconf]{IEEEtran}

\IEEEoverridecommandlockouts

\usepackage{balance}  

\usepackage{amsmath}
\usepackage{amsfonts}
\usepackage{amssymb}
\usepackage{graphicx}
\usepackage{caption}
\usepackage{subcaption}
\usepackage{paralist}
\usepackage{comment}
\usepackage{color}
\usepackage{url}
\usepackage{xspace}

\usepackage{float}
\usepackage{listings}
\usepackage[usenames,dvipsnames,svgnames,table]{xcolor}
\usepackage{bibspacing}
\usepackage{multirow}

\usepackage[noend]{algpseudocode} 
\usepackage{algorithm}

\algblockdefx[Name]{Struct}{End}%
  [1][Unknown]{\textbf{struct} #1 }%
  {\textbf{end struct} }%

\algnewcommand{\LineComment}[1]{\Statex \(\triangleright\) #1}

\newcommand{\eg}{\emph{e.g.,}\xspace}
\newcommand{\ie}{\emph{i.e.,}\xspace}

\newcommand{\vspacebeforesec}{\vspace*{-0.01in}}
\newcommand{\vspaceaftersec}{\vspace*{-0.01in}}

\newcommand{\mdds}{\texttt{MDDS}\xspace}
\newcommand{\nbqclock}{NbQ-CLOCK\xspace}

\newcommand{\PARTITLE}[1]{\vspace{0.05in}\noindent\textbf{#1}:}

\begin{document}

\date{}

\title{
Ingestion, Indexing and Retrieval of \\ 
High-Velocity Multidimensional Sensor Data on a Single Node
}

\author{
\IEEEauthorblockN{Juan A. Colmenares, Reza Dorrigiv, and Daniel G. Waddington}
\IEEEauthorblockA{
Samsung Research America\\ 
Mountain View, CA, USA}
\thanks{\textit{Corresponding author}: J.A.~Colmenares~\texttt{<juan.col@samsung.com>}}
\thanks{
R.~Dorrigiv is now with Google and D.G.~Waddington with IBM Research Almaden. 
They contributed to this work while employed at Samsung Research America.
}
}

\maketitle

\begin{abstract}
Sources of multidimensional data are becoming more prevalent, partly due to the
rise of the Internet of Things (IoT), and with that the need to ingest and
analyze data streams at rates higher than before.  
Some industrial IoT applications require ingesting millions of records
per second, while processing queries on recently ingested and historical data. 
Unfortunately, existing database systems suited to multidimensional data
exhibit low per-node ingestion performance, and even if they can scale
horizontally in distributed settings, they require large number of nodes to meet
such ingest demands.
For this reason, in this paper we evaluate a single-node multidimensional data
store for high-velocity sensor data. 
Its design centers around a two-level indexing structure, wherein the global
index is an in-memory R*-tree and the local indices are serialized kd-trees.
This study is confined to records with numerical indexing fields and range
queries, and covers ingest throughput, query response time, and storage
footprint.
We show that the adopted design streamlines data ingestion and offers ingress
rates two orders of magnitude higher than those of a selection of open-source
database systems, namely Percona Server, SQLite, and Druid.
Our prototype also reports query response times comparable to or better than
those of Percona Server and Druid, and compares favorably in terms of storage
footprint. 
In addition, we evaluate a kd-tree partitioning based scheme for grouping
incoming streamed data records.
Compared to a random scheme, this scheme produces less overlap between groups of
streamed records, but contrary to what we expected, such reduced overlap does
not translate into better query performance. 
By contrast, the local indices prove much more beneficial to query performance.  
We believe the experience reported in this paper is valuable to practitioners
and researchers alike interested in building database systems for high-velocity
multidimensional data.

\end{abstract}

\section{Introduction} \label{sec:intro}

Today sensors, systems, and automated processes generate increasing volumes of
data.
Many of these data sources continuously produce \emph{multidimensional records},
each with various numerical fields containing measured values and timestamps.
Prime examples are mobile devices with positioning modules and various sensors,
recording time, location, 
orientation, and other variables.
Other examples are the electric grid, smart buildings, and data centers.

With the proliferation of sensor deployments, multidimensional data streams are
expected to grow in number, but also be generated at higher rates.
Particularly, some Internet of Things (IoT) applications in industrial settings
(\eg power distribution telemetry~\cite{andersen2015}) require ingesting
unbounded streams each carrying tens of millions of records per second, while
running queries on recently ingested and historical data.
These demands are beyond the capabilities of most existing database systems and
have motivated the development of new ones. 
New time series databases~\cite{gorilla2015,btrdb2016}, with time as the only or
primary indexing field, have been geared to sustain such high ingestion rates.
Also aiming at high-velocity data, new online analytical processing (OLAP)
systems have been proposed
lately~\cite{druid14,braun2015,ramnarayan2016,dehne2016,pedreira2016}. 
These OLAP systems 
are able to index records by multiple fields (multidimensional indexing), 
offer sub-second responses to analytical queries,
and 
are designed to scale \emph{horizontally}. 
Unfortunately, they exhibit low per-node ingestion performance, often far less
than 1 million records per second, and hence require large number of nodes
to meet the ingest demands mentioned above.
A similar observation has been made in~\cite{streambox2017} about stream
processing engines.

For that reason, in this paper we go back to basics and revisit the ingestion,
indexing, storage, and retrieval of high-velocity multidimensional sensor data
on a single node.
We focus on records with numerical indexing fields and range queries, both
commonly used in sensor and system monitoring applications.
The main question we seek to address is 
\emph{whether it is possible to build a single-node multidimensional data store
able to sustain ingestion rates much higher than those of individual nodes of
existing database systems, while still offering similar query performance}.
To answer this question, we adopt a design, presented in \S\ref{sec:approach},
which centers around a hierarchical (two-level) indexing structure, reminiscent
of \mbox{EMINC}~\cite{zhang2009}, wherein the global index is an in-memory
R*-tree~\cite{beckmann1990,sellis1997} and the local indices are serialized
k-dimensional trees (kd-trees)~\cite{bentley1975}.
Based on this design, we built a multidimensional data store prototype, called
\mdds and described in \S\ref{sec:design}.
It implements a fast data ingestion path that executes almost entirely in
memory, leverages task parallelism, and makes data available to queries before
being persisted.
\mdds is conceived as a building block of the nodes in the data ingestion tier
of a distributed stream processing system, described in
\S\ref{sec:distsys-arch}.

Our results (\S\ref{sec:eval}) indicate that we can positively answer the
question. 
We observe that \mdds can sustain ingestion rates two orders of magnitude higher
than those of the selected competing systems (Percona
Server~\cite{percona-server}, SQLite3~\cite{sqlite}, and Druid~\cite{druid14}).
\mdds also reports query response times comparable to or better than those of
Percona Server (using a single multi-column index) and Druid, even though
\mdds's data retrieval path is not optimized. 
Moreover, \mdds, using no compression technique, compares favorably in terms of
storage footprint against the other systems.

In addition, we evaluate two methods for partitioning ingress streamed data,
referred to as \emph{data segmentation schemes} and presented in
\S\ref{sec:dataseg}. 
One scheme simply groups incoming records uniformly at random, while the other
does so by creating an in-memory kd-tree with the records and partitioning the
tree based on the number of nodes in subtrees. 
\mdds implements both schemes as part of its ingestion path. 
Compared to the random scheme, the kd-tree partitioning based scheme
produces less overlap between groups of streamed data records (\ie data
segments) at the expense of lower ingestion performance. 
We hypothesize that such reduced overlap among data segments improves query
response times by decreasing read amplification. 
Our results, however, \emph{do not} support this hypothesis; 
we observe that the kd-tree partitioning based scheme yields insignificant or no
reduction in query response time.
By contrast, the use of local indices, implemented as serialized kd-trees, is
much more beneficial to query performance.

The contributions of this paper are the following:
\begin{compactenum}
\item
An experimental study of the design of a single-node multidimensional data store
system for high-velocity sensor data. 
The study covers ingest throughput, query response time, and storage footprint.

\item
The characterization of two data segmentation schemes -- 
the random assignment scheme (\S\ref{sec:random-segmentation}) and 
the kd-tree partitioning based scheme (\S\ref{sec:kdtree-segmentation}) -- 
in terms of ingestion and query performance, including the effect query
processing has on data ingestion.

\item
A comparison of the \mdds prototype against a selection of well-known
open-source database systems: 
\emph{Percona Server}~\cite{percona-server}, with several storage engines
(XtraDB, MyISAM, and TokuDB), and \emph{SQLite3}~\cite{sqlite}, as
representatives of relational database management systems (RDBMSs), 
and \emph{Druid}~\cite{druid14}, a NoSQL OLAP system.

\end{compactenum}
 
In our evaluation, we use two publicly available datasets: 
\emph{NYC Taxi Trip} dataset~\cite{nyctaxi-data} and  
\emph{Global Historical Climatology Network - Daily} (GHCN-Daily)
dataset~\cite{ghcn-dataset,ghcn2012}. 
We also use two sets of custom SQL range queries, one for each dataset, which
are listed in Appendix~\ref{sec:test-queries}.

\section{Target Multi-Node System} \label{sec:distsys-arch}

\begin{figure}[!tbp]
\centering
\includegraphics[width=0.60\columnwidth]{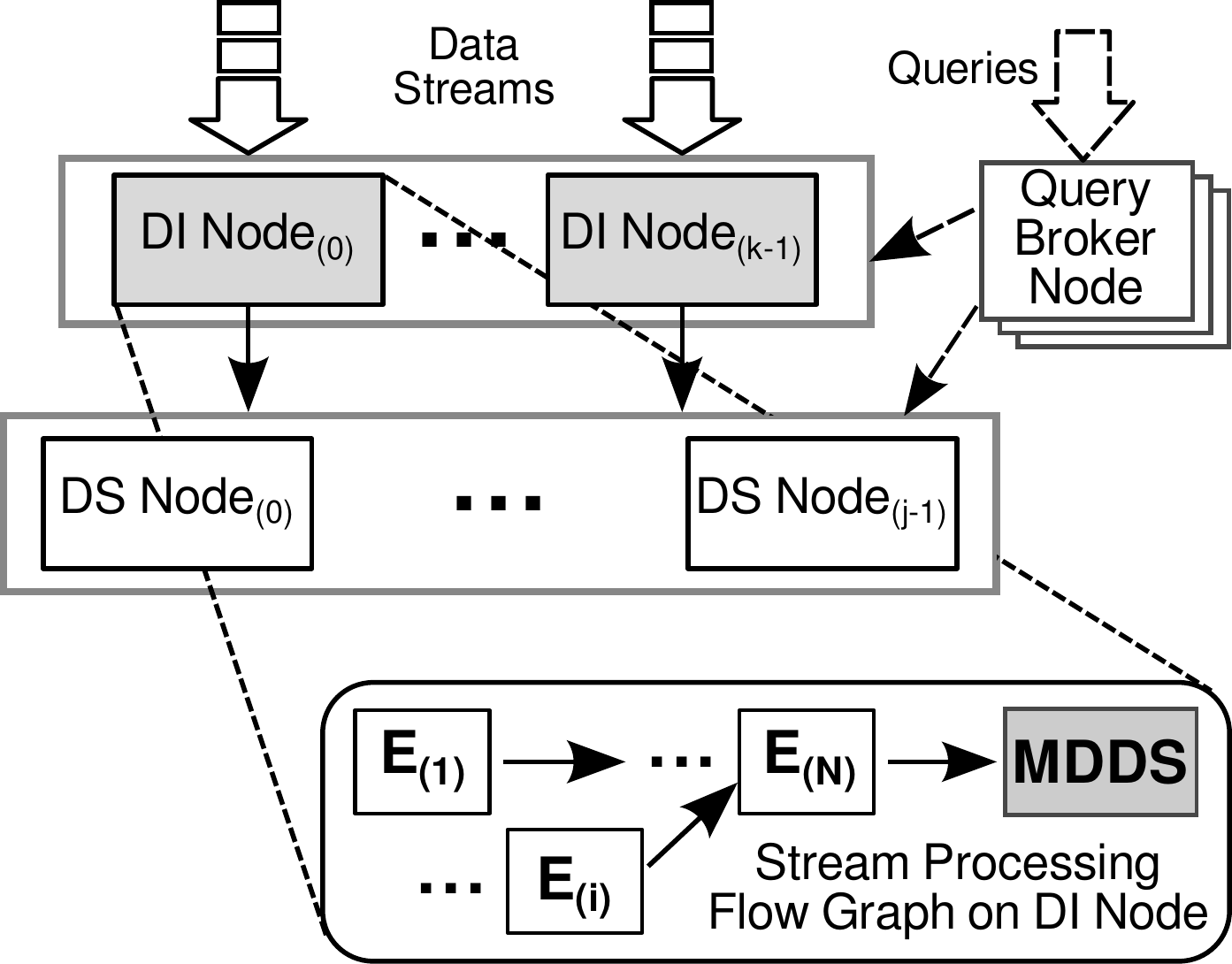}
\caption{
\small
Distributed data stream processing system with separate nodes
for data ingestion (DI), data storage (DS), and query brokering.
A multidimensional data store (MDDS) is a terminal vertex in the stream
processing flow graph on a data ingestion node.
}
\label{fig:dist-sys-design}
\vspace*{-0.15in}
\end{figure}

We consider a multi-node data stream processing system, whose basic architecture
is shown in Figure~\ref{fig:dist-sys-design}.
Similar to Druid~\cite{druid14}, the system includes separate nodes for data
ingestion, data storage, and query brokering. 
\emph{Data ingestion (DI) nodes} receive data streams and may perform some
computation on the data (\eg filtering and summarization).
They offer interim storage for raw or processed streamed data 
that have recently entered the system.
By contrast, \emph{data storage (DS) nodes} are responsible for permanently
storing data. 
DI nodes offload the data to DS nodes once a time interval expires, after
a data volume threshold is exceeded, or at times of light system's loads.
DI and DS nodes both serve queries on the data they possess. 
Moreover, \emph{query broker nodes} receive client queries and re-direct them to
the relevant DI and DS nodes; they may also aggregate and filter the query
results, if needed.
The systems also includes other nodes (not shown) in charge of activities such
as global data indexing, replication, and load balancing.

DI nodes are tailored to ingesting and storing append-only data at high velocity
(a write-heavy workload) while providing good query performance.  
They should then include data store subsystems featuring indexing structures
that are quick to populate and effective for query processing. 
We focus in this paper on one of such subsystems: a \emph{multidimensional data
store} that, as depicted in Figure~\ref{fig:dist-sys-design}, is meant to reside
in DI nodes as a terminal vertex in the stream processing flow graph. 
We envision the use of a single-node stream processing engine like
StreamBox~\cite{streambox2017}, which has been designed to maximize streaming
throughput and minimize latency on modern multicore hardware.

\section{Design Approach} \label{sec:approach}

The adopted approach to designing a single-node multidimensional data store
treats each incoming data stream as independent, and assumes that data records
in a given stream have the same structure. 
A \emph{multidimensional data record} is assumed to be a tuple 
$\mathbf{r} \triangleq \langle f_{1}, f_{2}, \dots, f_{N-1}, f_{N} \rangle$ 
with $N \geq 2$ fields, where two or more fields are of \emph{numerical} type
(integer, fixed-point, and floating-point).
From those numerical fields in the records, at least two are used for data
indexing, and we call them \emph{indexing dimensions}.

\begin{figure}[!tbp]
\centering
\includegraphics[width=1.0\columnwidth]{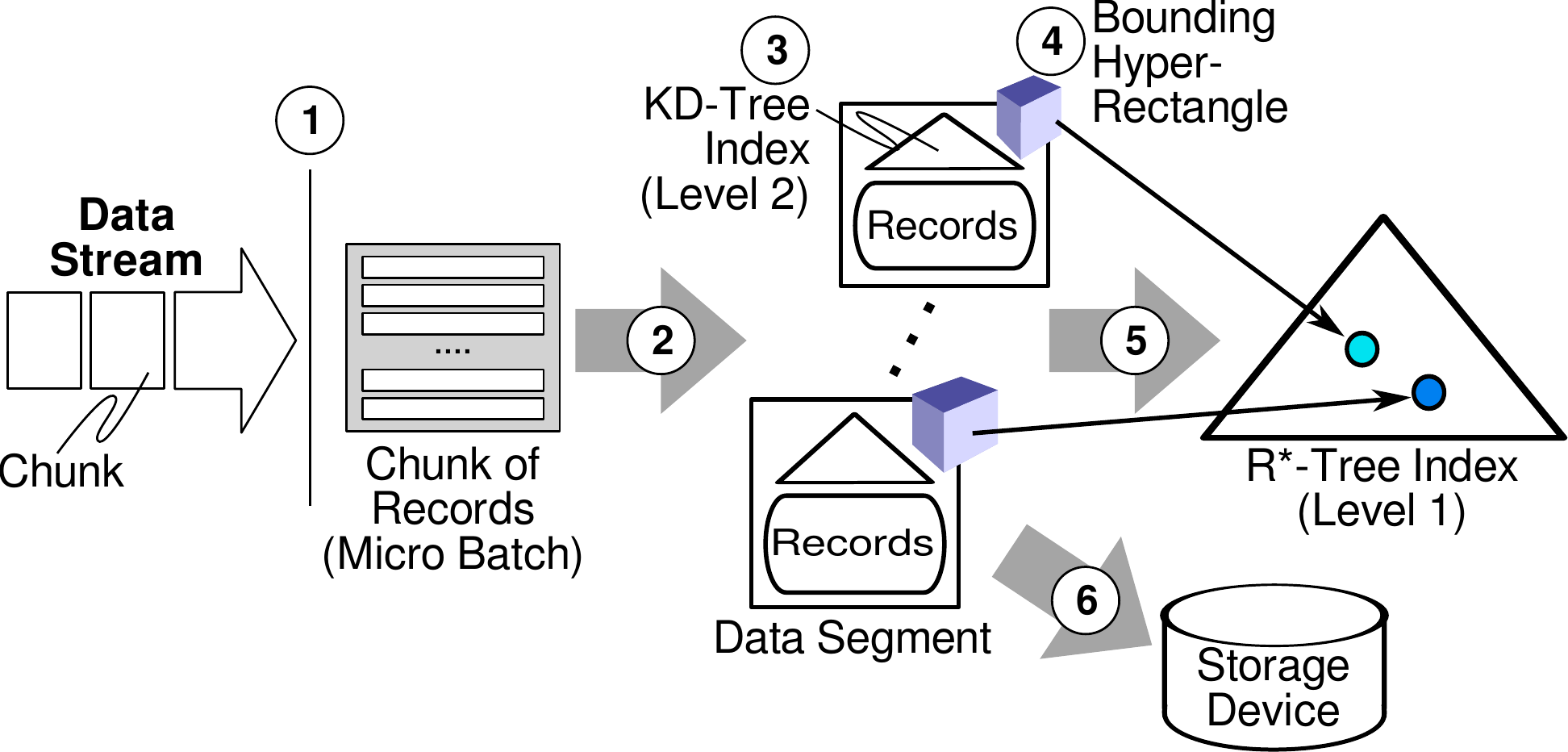}
\caption{
\small
Multidimensional data ingestion procedure.
The steps are as follows: 
(1) get a chunk of records, 
(2) divide the records in the chunk into data segments,
(3) build the second-level kd-tree index for each data segment,
(4) determine the bounding hyperrectangle for each data segment,
(5) insert the bounding hyper-rectangles into the first-level R*-tree index, and 
(6) write the data segments to permanent storage.
}
\label{fig:ingest-approach}
\vspace*{-0.15in}
\end{figure}

The data store includes a data ingestion path and a data retrieval path. 
The ingestion path, implemented by the \emph{ingestor} component, receives
chunks (micro batches) of streamed data and divides them into basic units of
storage called \emph{data segments}.
Each data segment contains multidimensional data records and a
kd-tree~\cite{bentley1975} to index them.
In addition, a data segment includes the dimensional ranges (the minimum and
maximum values per dimension) that encompass its data records and define the
segment's \emph{bounding hyperrectangle}.

Right after creating the data segments from a chunk, the ingestor inserts
references to the segments into an in-memory
R*-tree~\cite{beckmann1990,sellis1997}, which 
indexes the segments according to their hyperrectangles and makes
them promptly available to queries.
Data segments are accessible to queries from memory before becoming persistent,
and they are written to storage asynchronously.
Once written, data segments are served from the storage device (unless they are
in cache due to prior queries).
The steps the ingestor takes are illustrated in
Figure~\ref{fig:ingest-approach}. 

\begin{figure}[!tbp]
\centering
\includegraphics[width=0.45\columnwidth]{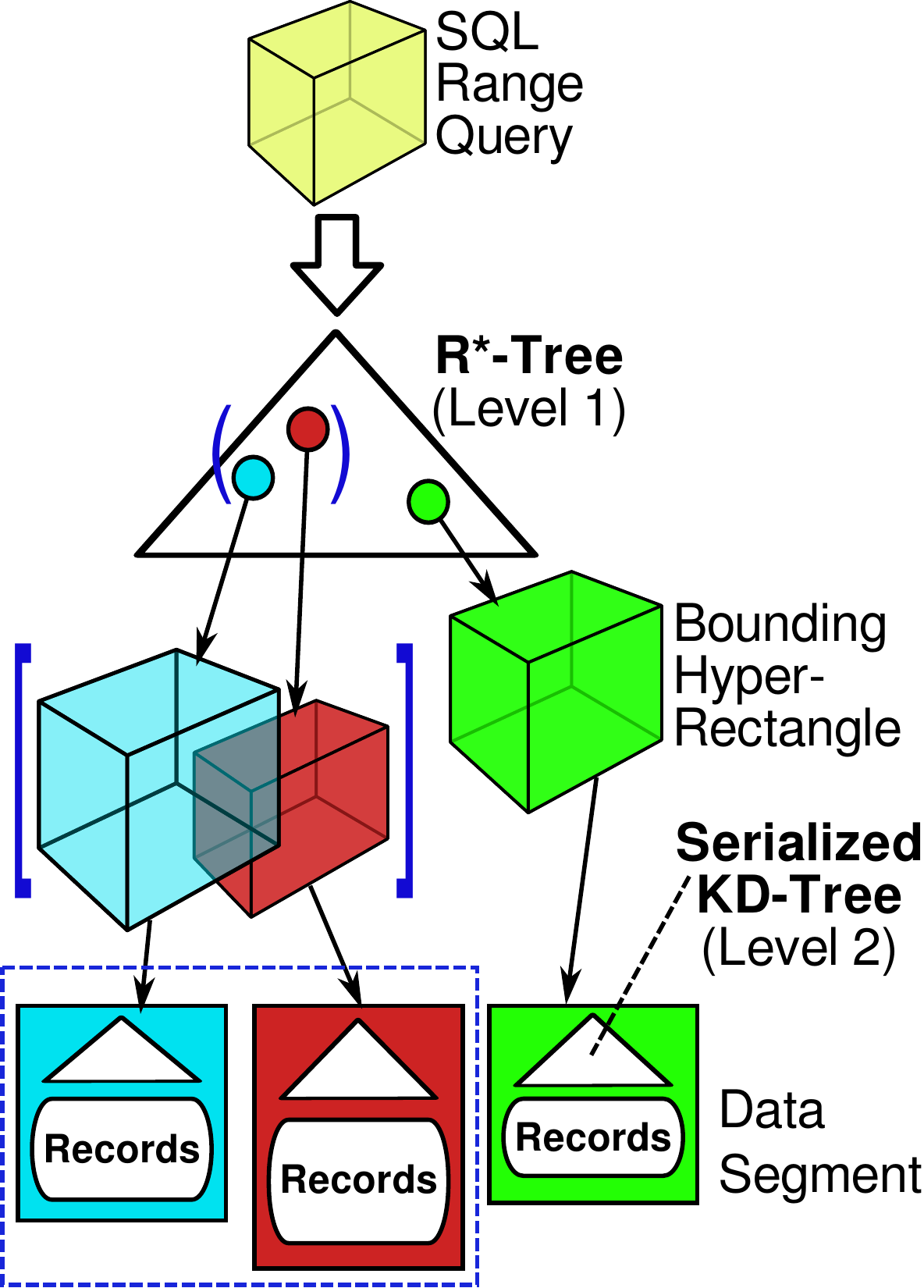}
\caption{
\small
Range query processing.
}
\label{fig:query-processing}
\vspace*{-0.15in}
\end{figure}

The data retrieval path, on the other hand, is implemented by a \emph{query
processor} component, which receives ad-hoc range queries and returns data
records within the specified ranges. 
As shown in Figure~\ref{fig:query-processing}, upon receiving a query, the query
processor first searches the R*-tree index for data segments whose
hyperrectangles overlap the query's range (\ie segments potentially with records
in the range).
Then, it reads the selected segments into memory (if not in cache),
inspects them, and returns the records in the range.

The adopted approach uses a \emph{two-level indexing structure}, reminiscent of
\mbox{EMINC}~\cite{zhang2009}. 
At the first level, the R*-tree indexes data segments based on their bounding
hyperrectangles, while at the second level, each data segment includes a kd-tree
to index the data records it contains. 
Also, under this approach, the ingestor can concurrently process multiple data
chunks while at the same time the query processor serves concurrent queries,
provided 
thread-safe access to the R*-tree is enforced.
Finally, the user is required to provide the structure of the data
records -- and implicitly the record size (in bytes).
The user must also indicate the indexing dimensions and supply the maximum
chunk size (as a record count) and the maximum segment size (in bytes).
These and other implementation details are discussed in \S\ref{sec:design}.

\section{Data Segmentation} 
\label{sec:dataseg}

As mentioned above, the ingestor component receives records in chunks and
performs a \emph{data segmentation} operation, which divides the records in each
chunk into groups to create the data segments.
Since data segmentation is key to sustaining high ingestion rates, it is
performed only in memory. 
Besides ingestion speed, it is desirable that data segmentation arrange the data
in a way that does not hinder good query performance. 
Next, we present two data segmentation schemes, and 
evaluate their merits later in \S\ref{sec:eval}.

\subsection{Uniformly Random Scheme} 
\label{sec:random-segmentation}

This scheme iterates over the data records in a chunk and assigns each record to
a data segment chosen uniformly at random from a given set.
The set initially contains empty data segments, and its cardinality is 
$\lceil\frac{c \times r}{S}\rceil$ 
where 
$c$ is the record count in the chunk, 
$r$ the record size in bytes, 
and 
$S$ the maximum segment size also in bytes.
Simplicity and speed are this scheme's main attributes.

\subsection{Kd-tree Partitioning Based Scheme}
\label{sec:kdtree-segmentation}

\begin{figure}[!tp]
\centering
\includegraphics[width=0.55\columnwidth]{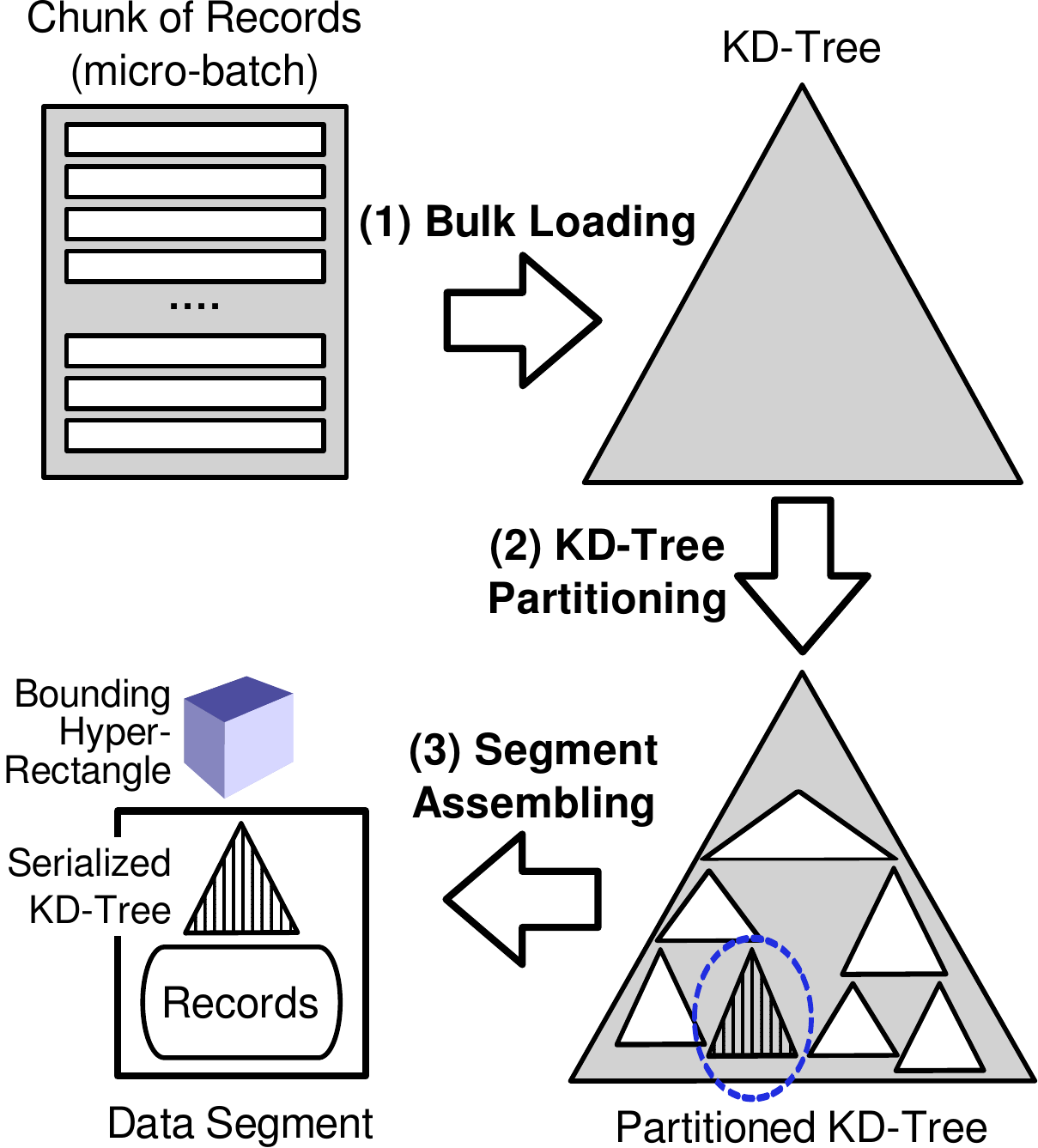}
\caption{
\small
Kd-tree partitioning based segmentation. 
}
\label{fig:chunk-segmentation-v2}
\vspace*{-0.20in}
\end{figure}

This more elaborate scheme tries to create well-populated data segments with
small overlap among their bounding hyperrectangles in order to limit read
amplification for query processing -- a property is often sought in distributed
data stores.
The scheme uses an in-memory kd-tree as central data structure and includes two main
steps -- bulkloading and partitioning -- that the ingestor performs upon the
arrival of each data chunk.
These two steps are described below and illustrated in
Figure~\ref{fig:chunk-segmentation-v2}.

\PARTITLE{Bulkload the records into a kd-tree} 
The ingestor first inserts all the records in the chunk into an empty kd-tree. 
Record bulkloading is performed recursively, where the indexing dimension at
each branching level is chosen in a round-robin fashion~\cite{ullman1988}.
In addition to links to the children, each node in the tree includes 
the number of nodes in its subtree (including itself). 
 
\PARTITLE{Partition the kd-tree} 
The ingestor traverses the kd-tree in depth-first pre-order and groups the
records based on the number of nodes in the subtrees.
It visits the subtree with higher node count first, and breaks ties by
continuing to the left subtree.
The traversal proceeds recursively until the ingestor finds a subtree with a
node count less than or equal to a predetermined \emph{maximum number of records
per segment} ($rps^{max}$).
In that case, the ingestor 
(1)~assigns the records in the subtree to a data segment from a given set, 
(2)~detaches the subtree from the kd-tree to avoid traversing it again,
and 
(3)~updates the subtree node count of ancestor nodes (still in the tree) as 
each recursive call returns.

$rps^{max}$ is calculated as $\lceil S / r \rceil \times \omega$, 
where 
$S$ is the maximum segment size, 
$r$ is the record size (with $r \ll S$), and 
$\omega$ is the \emph{overpacking factor}. 
By allowing the insertion of a greater number of records per segment, $\omega$
helps us compensate for the fact that this segmentation scheme tends to produce
partially filled data segments.
In our experience, $\omega = 4$ yields good results.

In addition to kd-tree bulkloading and partitioning, the ingestor performs a
third step: \emph{segment assembling} 
(see Figure~\ref{fig:chunk-segmentation-v2}).
This step is common to both schemes. 
It involves creating and populating the segments with the records assigned to
them as well as populating their bounding hyperrectangles and serialized
kd-trees.
Moreover, in both schemes the ingestor determines the dimensional ranges of each
segment's hyperrectangle as it assigns records to the segments.

\vspacebeforesec
\section{System Prototype} \label{sec:design}
\vspaceaftersec

This section presents our multidimensional data store prototype, called \mdds,
which implements the design discussed in \S\ref{sec:approach} and the segmentation
schemes of \S\ref{sec:dataseg}.

\subsection{Record Descriptor} \label{sec:rec-desc}

\mdds requires the user to provide a \emph{record descriptor file}, in XML
format, with the record structure of the ingress data stream.
Listing~\ref{code:nyc-rec-desc} presents the record descriptor for NYC taxi trip
data~\cite{nyctaxi-data} (with time fields in epoch instead of UTC). 
The file includes the UUID identifying the record's type and specifies the
record's fields (in order), their types, and given names. 
It also indicates the subset of numerical fields used as indexing dimensions and
the order in which they must be considered. 

\lstset{
language=xml,
tabsize=1,
caption={\small Record descriptor for NYC taxi trip data.},
label={code:nyc-rec-desc},
frame=single,
rulesepcolor=\color{gray},
framexleftmargin=1pt,
stringstyle=\color{blue},
keepspaces=true,
breaklines=false,
showstringspaces=false,
basicstyle=\scriptsize\ttfamily, 
emph={struct,indexing,dimensions},emphstyle={\bfseries},
}

\begin{lstlisting}[float,floatplacement=tbp]
<?xml version="1.0"?>
<description typeid="..."> <!-- UUID -->
<!-- The record format -->
<struct>
 <field name="medallion" type="char" array_len=33/>
 <field name="license" type="char" array_len=33/>
 <field name="vendor_id" type="char" array_len=4/>
 <field name="rate_code" type="int64_t"/>
 <field name="strfwd_flag" type="char" array_len=2/>
 <field name="pickup_datetime" type="epoch_t"/>
 <field name="dropoff_datetime" type="epoch_t"/>
 <field name="passenger_count" type="int64_t"/>
 <field name="trip_time_in_secs" type="int64_t"/>
 <field name="trip_distance" type="float"/>
 <field name="pickup_longitude" type="float"/>
 <field name="pickup_latitude" type="float"/>
 <field name="dropoff_longitude" type="float"/>
 <field name="dropoff_latitude" type="float"/>
</struct>
<!-- Fields for indexing -->
<indexing-dimensions> 
 <field name="pickup_latitude"/>
 <field name="pickup_longitude"/>
 <field name="pickup_datetime"/>
 <field name="passenger_count"/>
 <field name="trip_time_in_secs"/>
</indexing-dimensions> 
</description>
\end{lstlisting}

\subsection{Multidimensional Data Segment} \label{sec:dseg}

\begin{figure}[!tbp]
\centering
\includegraphics[width=0.55\columnwidth]{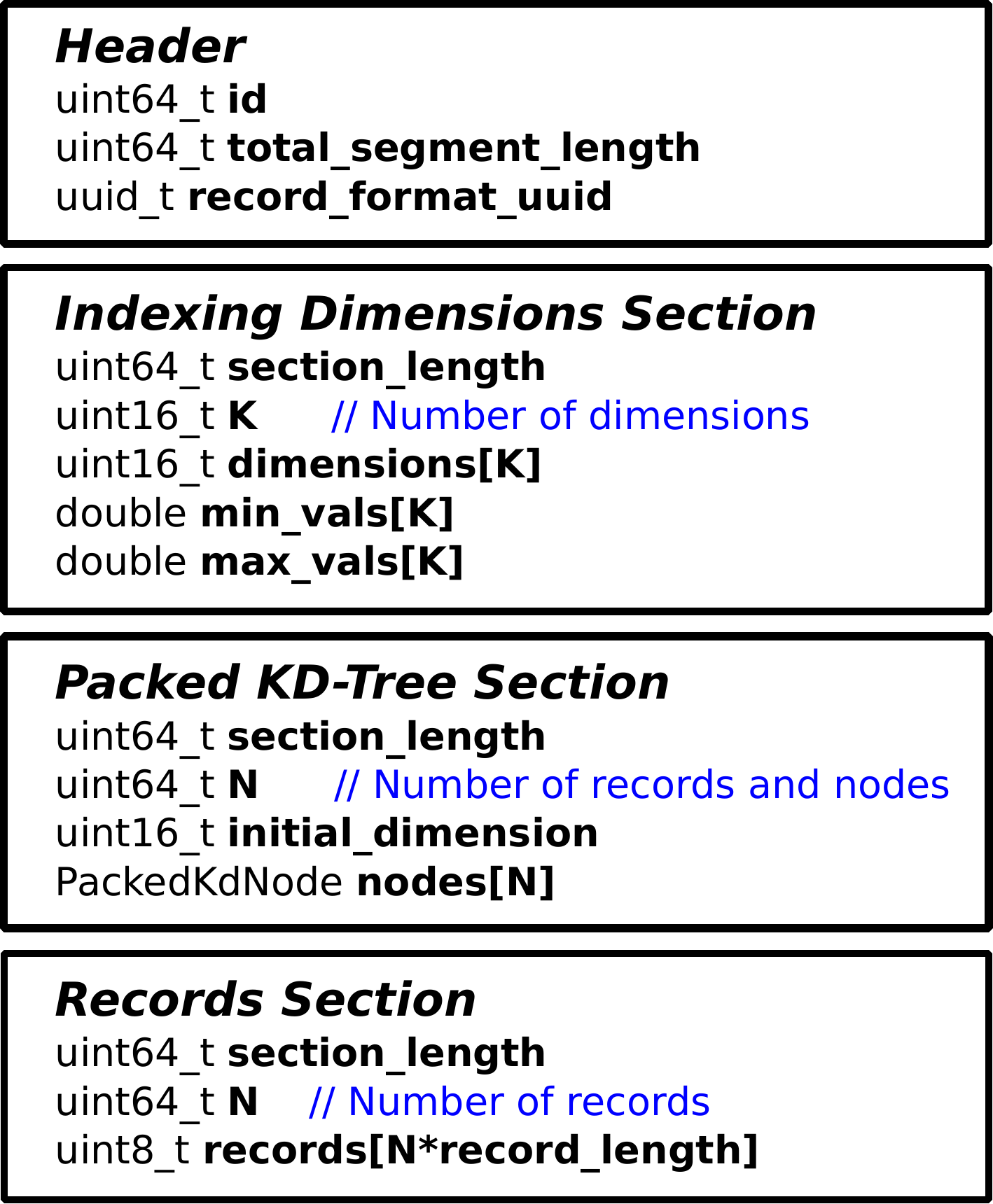}
\caption{
\small
Layout of multidimensional data segment.
}
\label{fig:md-dseg}
\vspace*{-0.15in}
\end{figure}

Figure~\ref{fig:md-dseg} shows the serialization layout of a data segment. 
It is the primary agreement on data representation between the data ingestor and
the query processor --- the former produces and stores data segments in this
format, while the latter inspects them to answer queries.

A data segment includes the following sections. 
First, the \emph{header} stores the segment's unique identifier, its total
length in bytes, and the record format's UUID (the same as in
the record descriptor).
Second, the \emph{indexing dimensions section} indicates the record fields used for
data indexing, as in the record descriptor. 
It also contains the minimum and maximum values per dimension that
define the bounding hyperrectangle for the records in the segment. 
Third, the \emph{packed kd-tree section} contains a kd-tree, in the form of an array of
nodes, that indexes the records in the segment.
Lastly, the \emph{records section} stores the sequence of data records as an
array of bytes.
These sections, except the header, are of variable length, which is 
stored in their \texttt{section\_length} field. 

Since most sections are self-explanatory, we only provide further details on the
packed kd-tree section. 
We use the term ``packed'' to highlight that the kd-tree is serialized and
embedded in the data segment. 
The query processor uses this tree to search for records in the data segment as
it often benefits query performance (see \S\ref{sec:dseg-eval}).
The packed kd-tree is stored as an array of \texttt{PackedKdNode}s, each
containing the position of the associated record in the segment's records
section, and the positions in the array of its left and right child (with
negative values denoting nil child nodes).

An important field in the packed kd-tree section is \texttt{initial\_dimension}.
When the query processor traverses a segment's packed kd-tree to serve a query,
it assumes the indexing dimensions at each branching level follow a round-robin
order~\cite{ullman1988}.
This field stores the dimension for the query processor to start round-robin
accross the dimensions in tree traversals.
The random segmentation scheme (\S\ref{sec:random-segmentation}) simply sets the
field to zero.
But, in the case of the kd-tree partitioning based scheme
(\S\ref{sec:kdtree-segmentation}), the value given to
\texttt{initial\_dimension} is not fixed because the packed kd-tree is built
from a subtree of an in-memory kd-tree, and the branching
dimension for the subtree's root may vary.

\subsection{Software Architecture} \label{sec:mdds-arch}

\begin{figure}[!tbp]
\centering
\includegraphics[width=0.65\columnwidth]{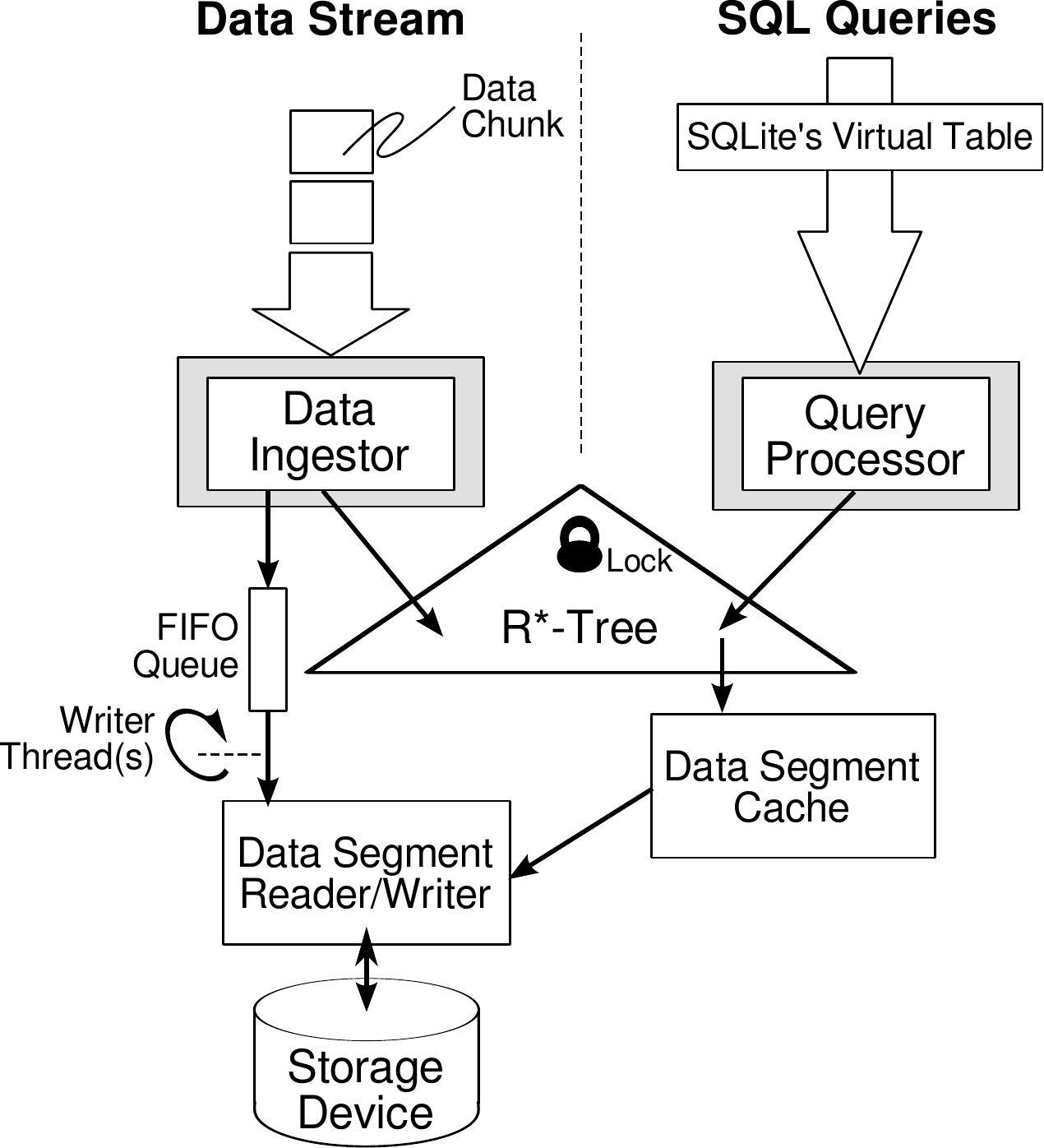}
\caption{
\small
\mdds architecture.
}
\label{fig:mdds-design}
\vspace*{-0.15in}
\end{figure}

Next, we discuss implementation details of \mdds's main components.
\mdds has been written in $\sim$8K lines of C++ code and its architecture is
illustrated in  Figure~\ref{fig:mdds-design}.

\PARTITLE{Data Ingestor}
The data ingestor produces data segments from every chunk of records it
receives.
To do so, it performs two operations on each chunk: 
\emph{data segmentation} and \emph{segment assembling}.
The former divides the records in the chunk into groups, each being the data
makeup of a separate segment, and the latter puts together the segments and
populates their sections. 
To ingest data at high velocity, the ingestor executes both operations solely in
memory.
It keeps the chunk's data records in their memory location throughout most of
the process and copies them at last when assembling the segments. 
Data segmentation, in particular, mainly relies on pointer manipulations and
numerical comparisons.

Since chunks can be processed independently, the ingestor can leverage task
(thread) parallelism to increase the ingestion rate.

\PARTITLE{Indexing and Writing Segments to Storage}
Right after being assembled, a data segment is in memory, in the right format to
be accessed by the query processor, and ready to be written to storage.
At this point, the ingestor creates a \emph{reference object} for the segment
and inserts the reference, along with the segment's bounding hyperrectangle,
into the R*-tree (step 5 in Figure~\ref{fig:ingest-approach}). 
Once its hyperrectangle and reference are in the R*-tree, the segment
becomes available to queries through the query processor.
Note that, as shown in Figure~\ref{fig:mdds-design}, a mutual exclusion lock
controls the access of ingestor threads and query processor threads to the
R*-tree, creating a potential contention point.

Next, the ingestor inserts (a pointer to) the reference into a lock-free FIFO
queue so that a \emph{writer thread} can asynchronously pull the reference from
the queue and write the segment to storage.  
After storing a segment, a write thread 
checks if the segment has been inserted in the \emph{data segment cache} (by the
query processor).
If the segment is in cache, the thread does not proceed and lets the
query processor be responsible for evicting and unloading the segment.
Otherwise, if the segment is not in cache, the thread marks the segment
as \emph{being-unloaded}, deletes it from memory (after ensuring its usage count
is zero), and then marks it as being \emph{not-in-memory}.
\mdds offers several configuration parameters for writer threads, such as 
thread count, activation period (in ms), and maximum number of
segments a thread can write every time it wakes up.

A segment's reference object is an in-memory representative of the segment.  
It maintains the segment's state, including 
its memory residency status,
whether it has been written to storage or not, and 
its current usage count in queries. 
Segment references allow the query processor to search for segments,  
independently of whether the segments are in memory or not.

Data segments are written to storage by the \emph{reader/writer component},
which provides a simple interface for other components in \mdds to read, write,
and delete segments. 
Its current POSIX implementation stores data segments as separate files in a
configurable directory.

\PARTITLE{Query Processor}
The query processor answers SQL range queries by performing the operations
illustrated in Figure~\ref{fig:query-processing}.  
It receives queries through a SQLite3's virtual table~\cite{sqlite-vtab},
which is automatically created from the record descriptor (\S\ref{sec:rec-desc}).
When a query arrives, the query processor first creates a \emph{cursor} object
that maintains the query's state.
It initializes the cursor by 
(1)~searching the R*-tree for references to data
segments whose bounding hyperrectangles overlap the requested range 
and 
(2)~inserting the references into a list inside the cursor.  
Once the cursor is initialized, the query processor uses it to iterate over the
identified segments and their records.

As the cursor traverses its list of data segments, it creates a \emph{record
iterator} object per segment.
The iterator encapsulates the logic to iterate over the records in the segment
and access those enclosed in the query's range. 
\mdds offers  two types of record iterators: 
a \emph{kd-tree iterator} that  uses the segment's packed kd-tree, 
and 
a \emph{sequential iterator} that ignores the packed kd-tree and simply accesses
the segment's records in sequence. 
Cursor objects can be configured to use either iterator type.

The query processor exploits temporal locality across queries by caching
data segments (rather than result sets).
It reserves a configurable amount of memory for the cache, and
evicts segments when the reserved memory is exhausted.
The segment cache uses the \emph{Non-blocking Queue-based CLOCK}
(\nbqclock)~\cite{eads2014}, a lock-free variant of the Generalized CLOCK
replacement algorithm~\cite{smith1978}.
We chose it due to its simplicity, good thread scalability, and fast update
operation.

Finally, the query processor can serve simultaneous queries.
This is enabled by processing queries in separate threads from a pool, and
having a cursor per query.
The thread pool's size is configurable.

\PARTITLE{Optimizations}
\mdds includes two optimizations for data ingestion. 
First, the ingestor uses a \emph{pre-allocated pool of kd-tree nodes} to
populate in-memory kd-trees.
Once a kd-tree is no longer needed, the ingestor returns the tree's nodes to the
pool.  
The pool is implemented atop a lock-free multi-producer/multi-consumer queue,
and used with both kd-tree partitioning based segmentation
(\S\ref{sec:kdtree-segmentation}) and random segmentation
(\S\ref{sec:random-segmentation}). 
With the latter, the ingestor still populates an in-memory kd-tree per segment
as an intermediate step to create the segment's packed kd-tree.
The pool is dimensioned at start time according to  
the maximum chunk size, 
record size, 
and maximum number of ingestor threads. 
This optimization was adopted after observing it performed better than Linux
default memory allocator and scalable memory allocators, such as
\texttt{tcmalloc}\footnote{http://goog-perftools.sourceforge.net/doc/tcmalloc.html}
and 
\texttt{jemalloc}.\footnote{http://jemalloc.net}

Second, to speed up the population of an in-memory kd-tree, the ingestor
\emph{does not} fully sort the records at each branching level.
Instead, at each branching level the ingestor -- \`a la quicksort -- selects a
record as the \emph{pivot} according to the indexing dimension being considered,
and relocates (the pointers to) the records so that records at the left (right)
of the pivot record have a dimension value lesser than or equal to (greater
than) the pivot's dimension value. 
The pivot is determined as the median of $M=3$ records selected uniformly at
random from the subset of records considered at current level.  $M$ is
configurable, but $M > 3$ offered no benefit.
For kd-tree partitioning based segmentation, the records are only sorted
at the root level because that tends to reduce the overlap between

\PARTITLE{Limitations}
\mdds has some limitations, mostly for simplicity of implementation. 
First, it can only ingest a single data stream. 
Second, \mdds stores the records as rows in the data segments; the advantages of
storing information as columns rather than rows are well
documented~\cite{abadi2008}, particularly for aggregation operations.
Despite these shortcomings, \mdds allows us to assess the merits of the
design approach in \S\ref{sec:approach} and evaluate the data segmentation
schemes in \S\ref{sec:dataseg}.

\vspacebeforesec
\section{Evaluation} \label{sec:eval}
\vspaceaftersec

In this section, we first characterize the two data segmentation schemes of
\S\ref{sec:dataseg} and study their influence on \mdds's ingestion and query
performance. 
We then evaluate \mdds against three well-known open-source database systems: 
Percona Server~\cite{percona-server}, 
SQLite3~\cite{sqlite}, and
Druid~\cite{druid14}.

We use two publicly available datasets. 
First, the \emph{NYC Taxi Trip} dataset~\cite{nyctaxi-data} 
with $\sim$169.67M records, 
each having 14 fields, 10 of them numerical and usable by \mdds as
indexing dimensions (see Listing~\ref{code:nyc-rec-desc}).
The time fields, originally in UTC format, were converted into epoch.
Second, the \emph{Global Historical Climatology Network - Daily} (GHCN-Daily)
dataset v.3.22~\cite{ghcn-dataset,ghcn2012}, 
from the US NOAA National Climatic Data Center. 
It contains daily climate observations from over 90K 
stations worldwide.
The GHCN-Daily dataset was converted into CSV flat records sorted by time.  
Each record has 7 fields, 6 of them numerical and indexable by \mdds 
(see Listing~\ref{code:ghcn-rec-desc}). 
We only use the first 100M records from a total over 700M. 

\lstset{
language=xml,
tabsize=1,
caption={\small Record descriptor for GHCN-Daily data.},
label={code:ghcn-rec-desc},
frame=single,
rulesepcolor=\color{gray},
framexleftmargin=1pt,
stringstyle=\color{blue},
keepspaces=true,
breaklines=false,
showstringspaces=false,
basicstyle=\scriptsize\ttfamily, 
emph={struct,indexing,dimensions},emphstyle={\bfseries},
}

\begin{lstlisting}[float,floatplacement=tbp]
<?xml version="1.0"?>
<description typeid="..."> <!-- UUID -->
<struct>
 <field name="time" type="epoch_t"/>
 <field name="station" type="char" array_len=12/>
 <field name="longitude" type="float"/>
 <field name="latitude" type="float"/>
 <field name="elevation" type="float"/>
 <field name="element_id" type="uint32_t"/>
 <field name="element_value" type="uint32_t"/>
</struct>
<indexing-dimensions> 
 <field name="element_id"/>
 <field name="latitude"/>
 <field name="longitude"/>
 <field name="elevation"/>
 <field name="element_value"/>
 <field name="time"/>
</indexing-dimensions> 
</description>
\end{lstlisting}

To evaluate query performance, we created two sets of range queries, one for
each dataset. 
For NYC Taxi data, we use 16 queries that cover 1km$\times$1km areas in New York
City and return the average passenger count on the selected taxi trips. 
Similar queries were used to evaluate Pyro~\cite{pyro2015}, a distributed
multidimensional data store. 
The queries were generated with ranges on two to five indexing fields, 
starting with pickup longitude and latitude, then incrementally adding pickup
time, passenger count, and trip time. 
The values for the ranges were chosen randomly, and out of numerous queries we
picked those that return non-zero values. 
For GHCN-Daily data, instead of random queries, we created 10 meaningful queries
(\eg \emph{the average snow depth for Mount McKinley (Alaska)}).
Our test queries along with table and index definitions are listed in
Appendix~\ref{sec:test-queries}.
Despite not being standard benchmarks, our tests are reproducible and serve well
the purpose of evaluating the segmentation schemes and comparing \mdds vs. other
systems. 


Our test platform is 
a Dell PowerEdge R720 server with 
two 2.50-GHz Intel Xeon CPU E5-2670 v2 processors 
(20 hardware threads with hyper-threading disabled), 
64GB of RAM, and 
an Intel 750 400GB SSD with ext4 file system.
It runs Ubuntu 14.04 LTS x86\_64 
(Linux kernel 3.13.0-71).

\subsection{Characterizing Segmentation Schemes} \label{sec:dseg-eval}

\begin{figure}[tp]
\centering
\includegraphics[width=0.80\columnwidth]{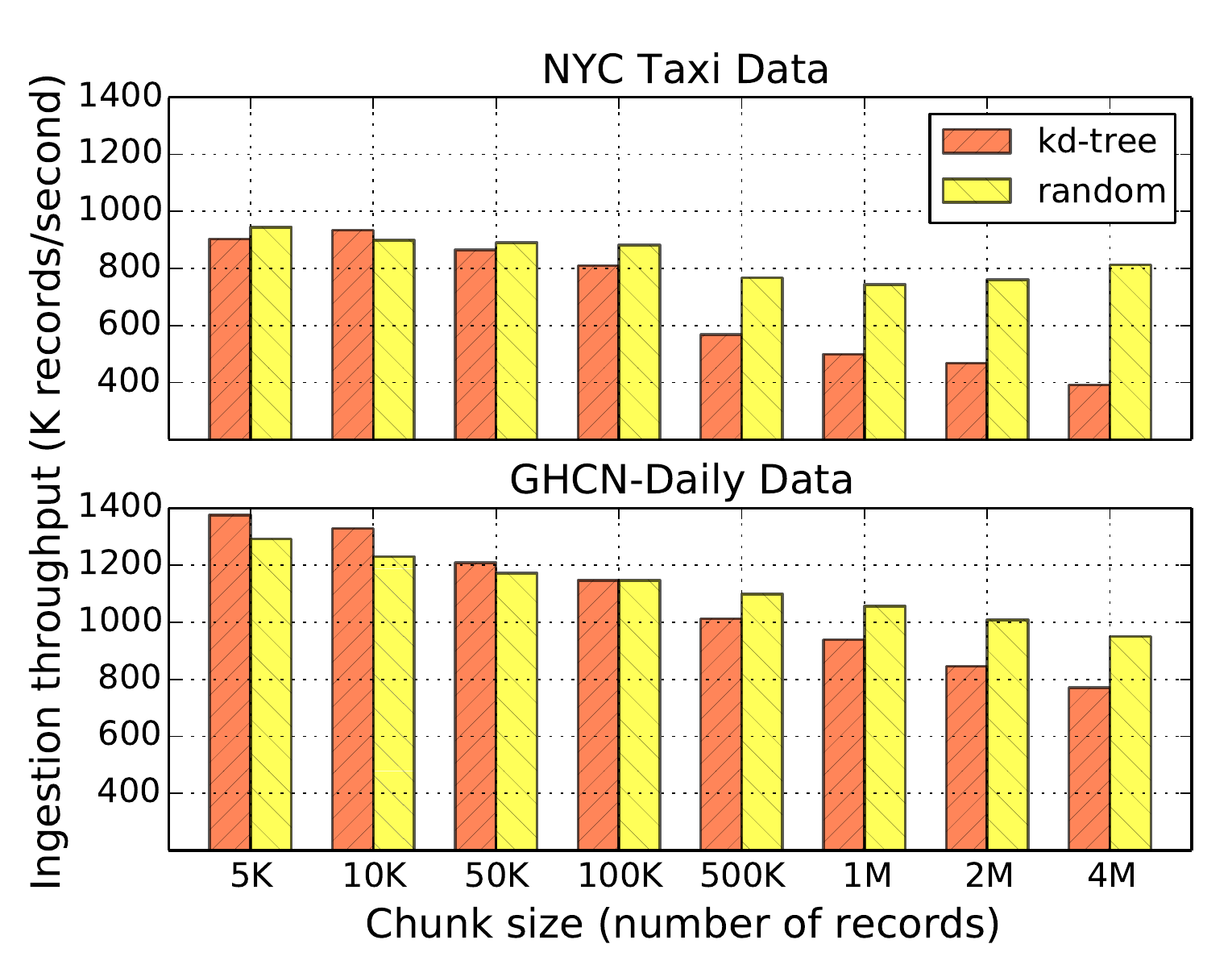}
\vspace*{-0.1in}
\caption{
Ingestion throughput for the random and kd-tree partitioning
segmentation schemes. 
}
\label{fig:data-seg-cmp}
\vspace*{-0.15in}
\end{figure}

We first evaluate \mdds's single-thread ingestion throughput with 
the random assignment scheme (\S\ref{sec:random-segmentation}) 
and
kd-tree partitioning scheme (\S\ref{sec:kdtree-segmentation}).
To isolate the effects of text-to-binary conversion and other auxiliary
operations, the input file with the test dataset in binary form is loaded
entirely into memory before being fed to \mdds. 
We exclude loading time from our results. 
We use five indexing fields for NYC Taxi data and six for GHCN-Daily data
(see Listings~\ref{code:nyc-rec-desc} and~\ref{code:ghcn-rec-desc}).

Figure~\ref{fig:data-seg-cmp} reports the mean throughput over 5 runs for
different maximum chunk sizes, ranging from 5K to 4M records, and a fixed target
segment size of 1MB.  
\mdds is able to sustain over 900K~records/s for NYC Taxi data and over
1.3M~records/s for GHCN-Daily data.
With both schemes, the throughput reaches the highest values with relatively
small chunks ($\leq$50K records), and decreases with the chunk size. 
Note that the kd-tree partitioning scheme has a more pronounced decrease in
throughput as the chunk size increases.
This is expected because kd-tree partitioning is more complex than random
assignment.
Similar results were obtained with 512KB and 4MB target segment sizes.

\begin{figure}[!tp]
\centering
\includegraphics[width=0.70\columnwidth]{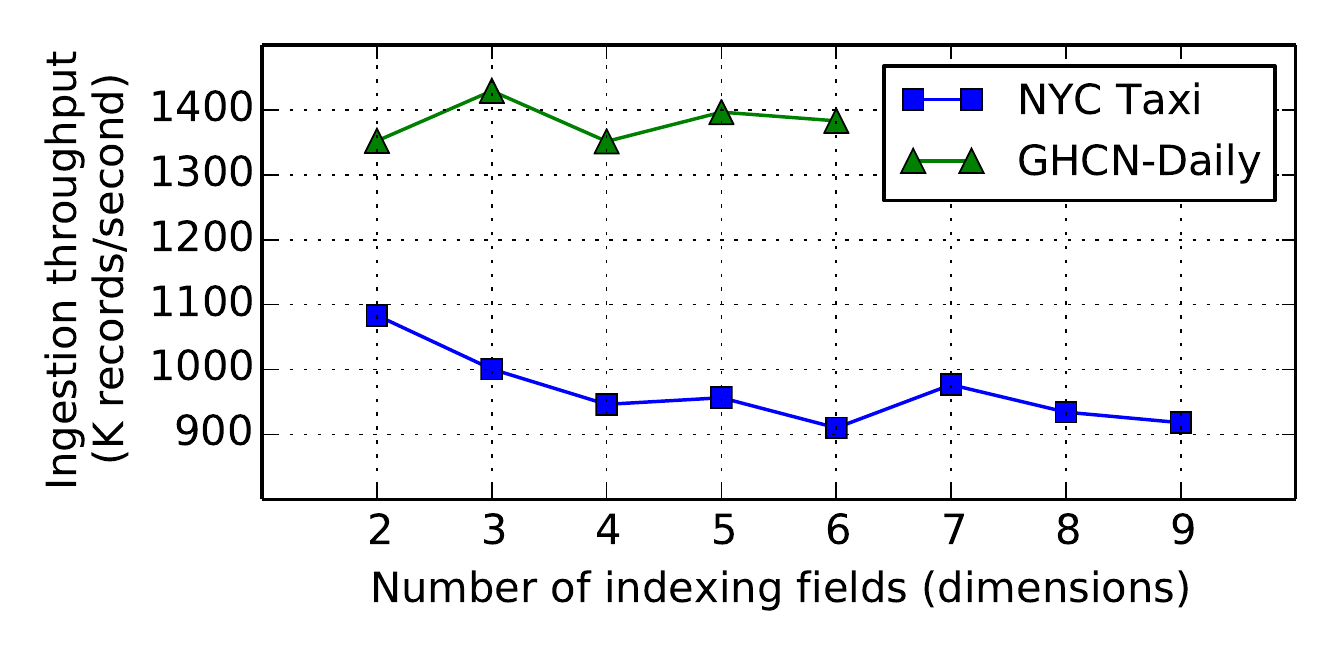}
\vspace*{-0.1in}
\caption{
Indexing dimensions vs. ingestion throughput for the kd-tree partitioning
scheme.
}
\label{fig:dims-cmp}
\vspace*{-0.15in}
\end{figure}

We also examine the impact of the number of indexing dimensions on \mdds's
ingestion throughput for the kd-tree partitioning scheme. 
(Dimension count has little to no influence for the random segmentation scheme.) 
Figure~\ref{fig:dims-cmp} reports the average throughput across 5 runs for
different dimension counts with both datasets; the maximum chunk size and target
segment size are both fixed to 10K records and 10MB, respectively.
We observe that increasing the dimension count may cause some degradation in
ingestion throughput (up to 20\% for NYC Taxi data). 
But, for small dimension counts ($\leq9$) the effect on ingestion performance is
small compared to that of chuck sizes.

\begin{figure*}[t!]
\centering
\begin{subfigure}[t]{0.5\textwidth}
\centering
\includegraphics[height=1.3in,width=3.7in]{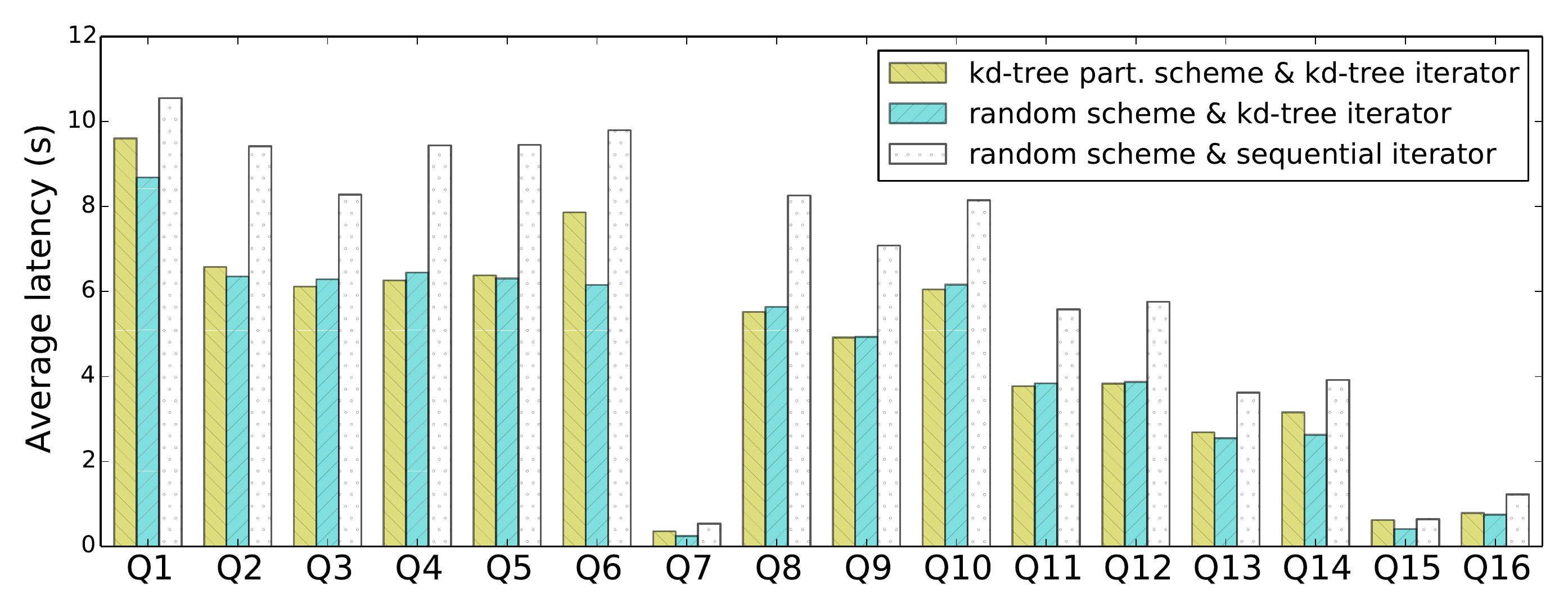}
\vspace*{-0.25in}
\caption{NYC Taxi data.}
\label{fig:nyctaxi-qperf-4-dataseg}
\end{subfigure}%
~ 
\begin{subfigure}[t]{0.5\textwidth}
\centering
\includegraphics[height=1.3in,width=2.8in]{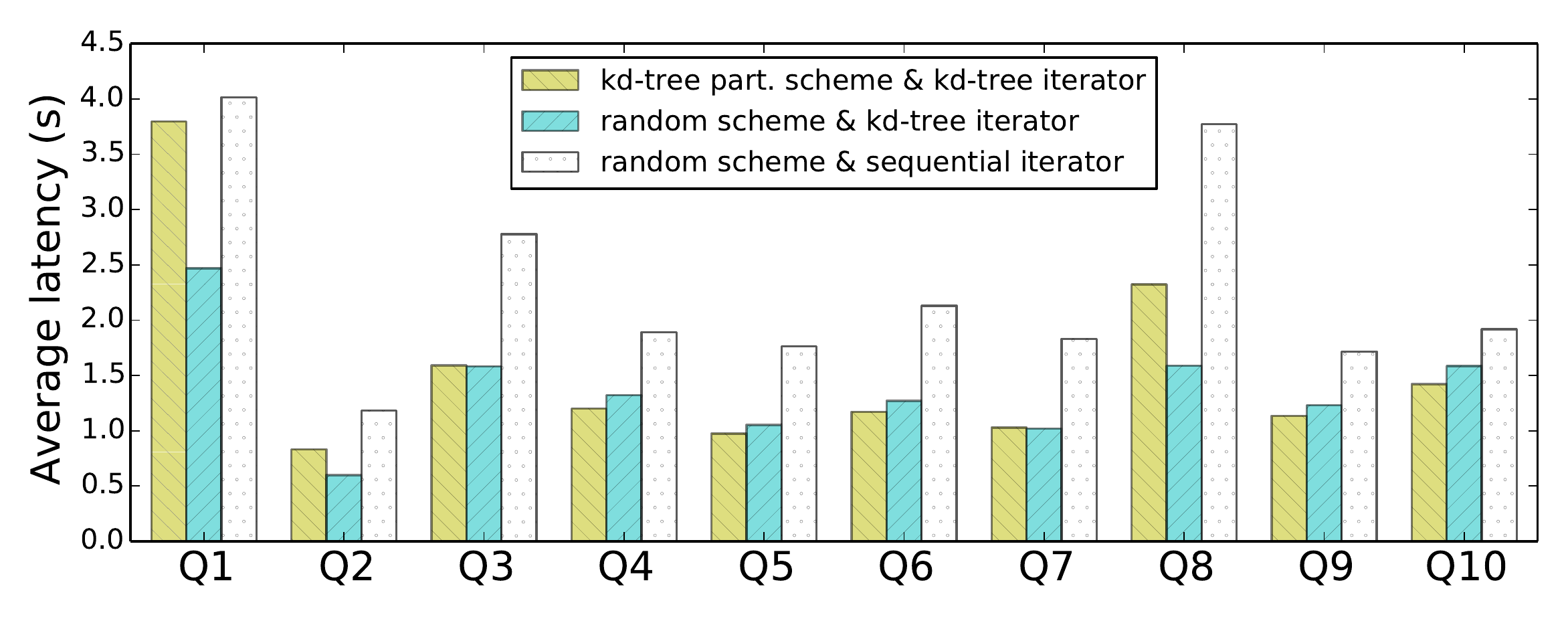}
\vspace*{-0.11in}
\caption{GHCN-Daily data.}
\label{fig:climate-qperf-4-dataseg}
\end{subfigure}
\vspace*{-0.10in}
\caption{
Influence of segmentation schemes and record iterators on \mdds's query
performance.
}
\label{fig:qperf-4-dataseg}
\end{figure*}

Next, we study the influence of both segmentation schemes on \mdds's query
performance.
We set \mdds with the same maximum chunk size and target segment size as before,
and a cache capacity of 1GB.
Figure~\ref{fig:qperf-4-dataseg} reports the average response times for our test
queries on both datasets over 5 runs, after a warm-up phase of 2 runs to avoid
performance variability due to cold cache.
We observe that compared to the random scheme, the kd-tree partitioning scheme
brings no significant reduction to response time and is even worse for some queries 
(\eg Q1 and Q6 in Figure~\ref{fig:nyctaxi-qperf-4-dataseg}, and Q1 and Q8 in
Figure~\ref{fig:climate-qperf-4-dataseg}). 
The reason is that the way the kd-tree partitioning scheme organizes the data
into segments either increases or just reduces a little the number of segments
that need to be inspected to answer the queries (\ie data segment selectivity
does not improve).

%
%
%
%

We should note, however, that the kd-tree partitioning scheme produces the
intended result: it tends to create data segments with much less overlap than
those created by the random scheme.
Table~\ref{table:seg-overlap} compares the overlap between segments created by
both segmentation schemes for two maximum chunk sizes (10K and 10M records) and
four target segment sizes (256KB, 512KB, 1MB, and 4MB).
The overpacking factor of kd-tree partitioning is adjusted so that both schemes
produce roughly the same number segments with similar average sizes.
We observe that, compared to the random scheme, the kd-tree partitioning scheme
offers a reduction in segment overlap of 2$\times$-51$\times$ in all cases but
one (GHCN-Daily, chunk size 10K, segment size 512KB), at the expense of extra
computational cost and much more variability in segment sizes
(26$\times$-100$\times$ increase in standard deviation).
Unfortunately, such reduction in segment overlap does not yield commensurate
reduction in latency for our test queries.

\begin{table}[!tbp]
\scriptsize
\centering
\begin{tabular}{|l|c|c|c|rr|}
\hline
\multirow{2}{*}{\textbf{Dataset}} & \multirow{2}{*}{\textbf{Scheme}} 
      & \textbf{Chunk}  & \textbf{Seg.} & \multicolumn{2}{c|}{\textbf{\# Segment Overlaps}} \\ 
      &                              
      & \textbf{Size}   & \textbf{Size}    & \textbf{Mean} & \textbf{(StdDev)} \\ 
\hline
\hline
\multirow{8}{*}{NYC Taxi}
              & random  & \multirow{4}{*}{10K} & \multirow{2}{*}{256KB} &  1,945.6 & (1,319.9) \\ 
              & kd-tree &                      &                        &   692.8  &  (677.5) \\ 
\cline{2-2} \cline{4-6}
              & random  &                      & \multirow{2}{*}{512KB} &  1,087.5 & (684.4) \\ 
              & kd-tree &                      &                        &    613.4 & (405.9) \\ 
\cline{2-6}
              & random  & \multirow{4}{*}{1M}  & \multirow{2}{*}{1MB}   &  1,230.6 & (1,097.7) \\ 
              & kd-tree &                      &                        &    24.3  &    (56.2) \\ 
\cline{2-2} \cline{4-6}
              & random  &                      & \multirow{2}{*}{4MB}   &    405.9 & (329.9) \\
              & kd-tree &                      &                        &     27.8 &  (52.4) \\ 
\hline
\hline
\multirow{8}{*}{GHCN-Daily}
              & random  & \multirow{4}{*}{10K} & \multirow{2}{*}{256KB} &  816.9  & (2,182.9) \\ 
              & kd-tree &                      &                        &  288.4  & (1,099.2) \\ 
\cline{2-2} \cline{4-6}
              & random  &                      & \multirow{2}{*}{512KB} &  416.1  & (1,111.5) \\ 
              & kd-tree &                      &                        &  396.5  & (1,061.7) \\ 
\cline{2-6}
              & random  & \multirow{4}{*}{1M}  & \multirow{2}{*}{1MB}   &  3,681.8 & (167.1) \\ 
              & kd-tree &                      &                        &    508.2 & (480.7) \\ 
\cline{2-2} \cline{4-6}
              & random  &                      & \multirow{2}{*}{4MB}   &  981.2 &  (39.4) \\
              & kd-tree &                      &                        &  255.9 & (196.9) \\ 
\hline
\end{tabular}
\caption{
Number of overlaps among segments for the random and kd-tree partitioning
segmentation schemes. 
}
\label{table:seg-overlap}
\end{table}


More important is the strong influence on query performance of the record
iterators discussed in \S\ref{sec:mdds-arch}. 
Figure~\ref{fig:qperf-4-dataseg} also compares \mdds's query latency with the
kd-tree iterator, which leverages the packed kd-tree in each data segment, and
the sequential iterator, on data ingested using the random segmentation scheme.
It is clear that packed kd-trees along with the kd-tree iterator offer important
(17\%-58\%) reductions in response time across all our test queries.




\subsection{System Performance Comparison} \label{sec:sys-eval}

\begin{table}[!tp]
\footnotesize
\centering
\begin{tabular}{lp{1.6in}}
\textbf{System} & \textbf{Storage Engine \& Subsystems } \\
\hline
\multirow{3}{1.3in}{Percona Server 5.6.27-76.0~(GPL)~\cite{percona-server}} 
    & XtraDB (enhanced InnoDB) \\  
    & MyISAM \\ 
    & TokuDB (fractal trees)~\cite{percona-tokudb} \\ 
\hline
SQLite 3.8.7.4~\cite{sqlite} & Default \\
\hline
\multirow{3}{1.3in}{Druid 0.9.0 stable~\cite{druid14,druid-src}} 
    & Real-time ingestion node. \\ 
    & Local deep storage, historical node, \\
    & and indexing service. \\ 
\hline
\end{tabular}
\caption{Database systems used for comparison.}
\label{table:other-systems}
\end{table}

Next, we compare the performance of \mdds and the open-source database systems
listed in Table~\ref{table:other-systems}.
First, 
\emph{Percona Server}~\cite{percona-server} is an
enterprise-ready RDBMS and a replacement for MySQL~\cite{mysql} 
with advertised superior performance. 
We evaluate its main storage engines: 
XtraDB (an enhanced version of MySQL's InnoDB), 
MyISAM, and 
TokuDB.
TokuDB~\cite{percona-tokudb}, in particular, uses a write-optimized index 
called \emph{fractal tree}~\cite{perconaft}, which is
essentially a B-tree augmented with per node buffers, and 
a variant of B$^{\epsilon}$-tree~\cite{brodal2003}. 
Compared to B-trees, fractal trees offer faster data ingestion with
matching or improved performance for 
range queries. 
Second, 
\emph{SQLite3}~\cite{sqlite}, a popular embedded relational database, is
evaluated because it is the basis of \mdds's query processor. 
Third,
\emph{Druid}~\cite{druid14,druid-src} is a NoSQL, column-oriented, distributed,
real-time analytical data store designed for business intelligence (OLAP)
queries on event data.
It has been adopted by companies such as 
Alibaba, Cisco, eBay, 
Netflix, and Yahoo, 
and production Druid clusters have scaled to over 3 million records per
second~\cite{about-druid}.

Percona Server and SQLite3 store NYC taxi data in a table and use a multi-column
index with the indexing fields in the same order as in
Listing~\ref{code:nyc-rec-desc}.
For Druid, the indexing fields are specified as \emph{dimensions} (in OLAP
parlance), with \texttt{passenger\_count} being also a \emph{metric} so its
aggregates can be calculated. 
Note that \texttt{pickup\_datetime} is a dimension and not the timestamp column. 
The reason is that the timestamp column only accepts current time values (within
a specified time window), and stale records are omitted; so, to avoid missing
records, the timestamp column is filled out with the current ingestion time.
For GHCN-Daily data, Percona Server and SQLite3 use a table with four
multi-column indices: three with two indexing fields and one with four.
These indices are tailored to our test queries, and unlike
Listing~\ref{code:ghcn-rec-desc}, they do not include the \texttt{time} field
since it appears in no query.
In Druid, as before, the indexing fields are dimensions, with
\texttt{element\_value} being also a metric and the timestamp column storing the
current ingestion time.

We run the systems on our test server (single node), where they store data on an
SSD. 
\mdds uses the \emph{random segmentation scheme} and \emph{kd-tree record
iterator} for better ingestion rates and query response times, as shown in
\S\ref{sec:dseg-eval}.
Similarly, Percona Server and its engines, SQLite3, and Druid are configured for
fast ingestion and good query performance.
The systems' configurations are summarized in
Table~\ref{table:sysconf-4-ingestion}; variables not shown use default values. 
Moreover, query results were verified across the systems to ensure correctness.

\begin{table}[!tb]
\footnotesize
\centering
\begin{tabular}{ll}
\textbf{System / Engine} & \textbf{Variables} \\
\hline
\multirow{4}{1.0in}{Percona Server\\(for the different engines)}
  & autocommit = 0 \\
  & foreign\_key\_checks = 0 \\
  & sql\_log\_bin = 0 \\
  & performance\_schema  =  0 \\
\hline
\multirow{5}{*}{XtraDB}  
  & innodb\_buffer\_pool\_size = 32G \\
  & innodb\_flush\_method = O\_DIRECT \\
  & innodb\_doublewrite = 0 \\
  & innodb\_flush\_log\_at\_trx\_commit = 2 \\
  & innodb\_log\_buffer\_size = 8M \\ 
\hline
MyISAM & ---                \\ 
\hline
\multirow{3}{*}{TokuDB}  
  & tokudb\_cache\_size = 48G \\
  & tokudb\_directio = 1 \\
  & tokudb\_loader\_memory\_size = 24G \\
\hline
\multirow{3}{1.0in}{SQLite\\(set via PRAGMA statements)}  
  & synchronous = OFF \\
  & journal\_mode = OFF \\
  & ignore\_check\_constraints = ON \\
\hline
Druid Real-Time Node & maxRowsInMemory = 500,000 (default) \\
\hline
\multirow{5}{1.0in}{Druid \& Tranquility Loader Application} 
  & maxRowsInMemory = 5,000,000 \\ 
  & Druid worker threads = 10  \\ 
  & HTTP (REST) threads = 10\\ 
  & windowPeriod = 1 min \\
  & intermediatePersistPeriod = 1 min \\
\hline
\multirow{4}{*}{\mdds}  
  &  seg. scheme = random assignment \\
  &  record iterator = kd-tree iterator \\
  &  target segment size = 1M \\
  &  overpacking factor = 4 \\
  &  writer thread period = 500 ms \\
\hline
\end{tabular}
\caption{
System configurations.  
}
\label{table:sysconf-4-ingestion}
\end{table}

\subsubsection{Single-Threaded Ingestion} 
\label{sec:1th-ingest-perf}


We evaluate \mdds's single-threaded ingestion against other systems' bulk
loading.\footnote{
We use bulk loading as it offers higher ingestion rates compared to insertion
statements.
}
Ingestion rates are measured with different maximum chunk sizes.
We load data into Percona Server and SQLite using a shell script that splits the
CSV data file into chunks and feeds them in FIFO order from \texttt{tmpfs} (RAM)
with the database's bulk load command.
To isolate effects of auxiliary operations, the ingestion throughput calculation
only includes the execution times of bulk load commands.
In the case of Druid, we load data through its \emph{Real-Time Node}; it uses a
local firehose~\cite{druid-firehose} to ingest records from a CSV file
(stream-pull ingestion), and dedicates a single thread to data consumption and
aggregation. 
We only report Druid's results with \texttt{maxRowsInMemory}~$=500,000$ (the
default value); this parameter indicates the number of rows to aggregate in
memory before persisting (similar to \mdds's maximum chunk size), but we
observed no significant variation in ingestion performance when experimenting
with it.\footnote{
We also omit the results obtained with Druid's Tranquility loader application,
which will be introduced later, because its single-threaded ingestion throughput
was 2-3$\times$ lower than that of Druid's Real-Time Node.
}  
For \mdds, each dataset is provided in two forms: a CSV file and a binary file. 
As in \S\ref{sec:dseg-eval}, the input file is loaded entirely into memory
before being fed to \mdds. 
Finally, we ensure across the experiment that each run uses a fresh system; \ie
before loading the data, the database is deleted, if it exists, and a new one is
created.

\begin{figure}[tp]
\centering
\includegraphics[width=0.9\columnwidth]{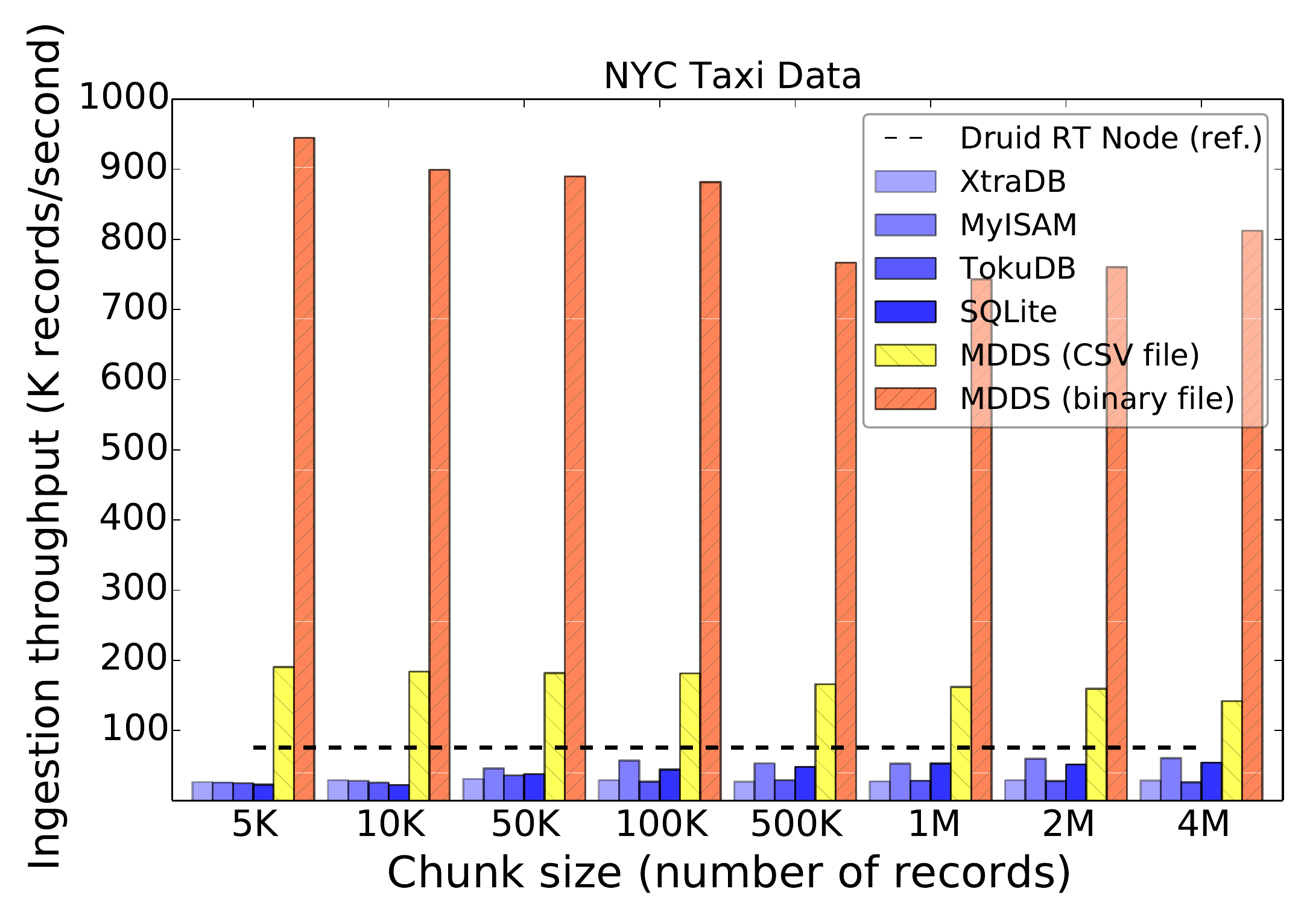}
\vspace*{-0.1in}
\caption{
Single-threaded ingestion (bulk load).
}
\label{fig:1t-bulkload-throughput}
\end{figure}

We observe that \mdds outperforms the other systems by at least 2$\times$ when reading
from a CSV file. 
Figure~\ref{fig:1t-bulkload-throughput} reports the systems' ingestion rates for
NYC Taxi data. 
Results for GHCN-Daily dataset are similar, but more favorable to \mdds; 
they are omitted for brevity's sake. 
Figure~\ref{fig:1t-bulkload-throughput} also shows that \mdds reaches much
higher ingestion throughput when data comes from a binary file rather than a CSV
file (\eg $\sim$4.5$\times$ higher for 10K-record chunks). 
We observe that the text-to-binary conversion (when ingesting CSV files) is the
main hotspot, eating a significant number of cycles and masking \mdds's actual
performance. 
As shown in Figure~\ref{fig:data-seg-cmp},  with data in binary form \mdds
reaches above 900K and 1.3M records/s for NYC Taxi and GHCN-Daily datasets,
respectively; \ie 11$\times$-16$\times$ better than Druid's Real-Time Node (the
best among the other systems).
Note that ingesting binary data is not uncommon, especially when \mdds is a
terminal vertex in a stream processing flow graph and up-stream vertices have
performed the necessary conversions (see Figure~\ref{fig:dist-sys-design}).


%
%
%
%
%
%
%

\subsubsection{Multi-Threaded Ingestion} \label{sec:xth-ingest-perf}

Here we compare the ingestion rates \mdds and the other systems can sustain with
multiple data feeder threads. 
All the systems, except Druid, load the data the same way. 
A \emph{data source component} reads the whole binary data file into a memory
region and divides the region into approximately even slices, one per feeder
thread. 
It then gives the slices to the threads that concurrently push the data in
chunks onto the system under test.

The data source component feeds data directly to \mdds, but in the case of
Percona Server and SQLite it does so through a \emph{wrapper component}.
The wrapper issues SQL \texttt{insert} statements, with the records received
from the data source component, to the database. 
It opens a separate database connection for each feeder thread and
configures the connections for high-ingestion throughput (\eg no autocommit).
For better performance, the wrapper uses prepared \texttt{insert}
statements bound to memory buffers; this way we avoid unnecessary data
conversions, but still data records must be copied into the buffers.
In addition, a transaction is committed for each data chunk.
In this experiment, Percona Server and SQLite are configured as
before (see Table~\ref{table:sysconf-4-ingestion}).

The data source component is configured to feed \mdds with chunks of up to 10K
records, which yield the second highest single-thread ingestion rate in
Figure~\ref{fig:1t-bulkload-throughput}.
For Percona Server and SQLite, the maximum chunk size is set to 1M records. 
We choose this value because with it Percona Server's engines and SQLite reach
ingestion throughputs close to their best for our datasets, and and we observe
no significant improvement with larger chunks.

In the case of Druid, we load data with a stand-alone Java application written atop
Tranquility~\cite{tranquility}, a client library for sending real-time event
streams directly to Druid (stream-push ingestion) without involving
any real-time node.
The application runs a configurable number of data loader threads. 
Instead of binary data, the threads load CSV data as it matches Traquility's
Tranquilizer API, avoiding unnecessary data conversion.
The threads feed data in chunks to prevent out of memory errors and other low memory
issues in the application and Druid's subsystems 
(\eg indexing service and historical node).
Through experimentation, we observe the loader application performs best with
chunks of 1M records and the configuration variables in
Table~\ref{table:sysconf-4-ingestion}.

As before, we ensure each run uses a fresh system.

%
%

\begin{figure}[tp]
\centering
\includegraphics[width=0.95\columnwidth]{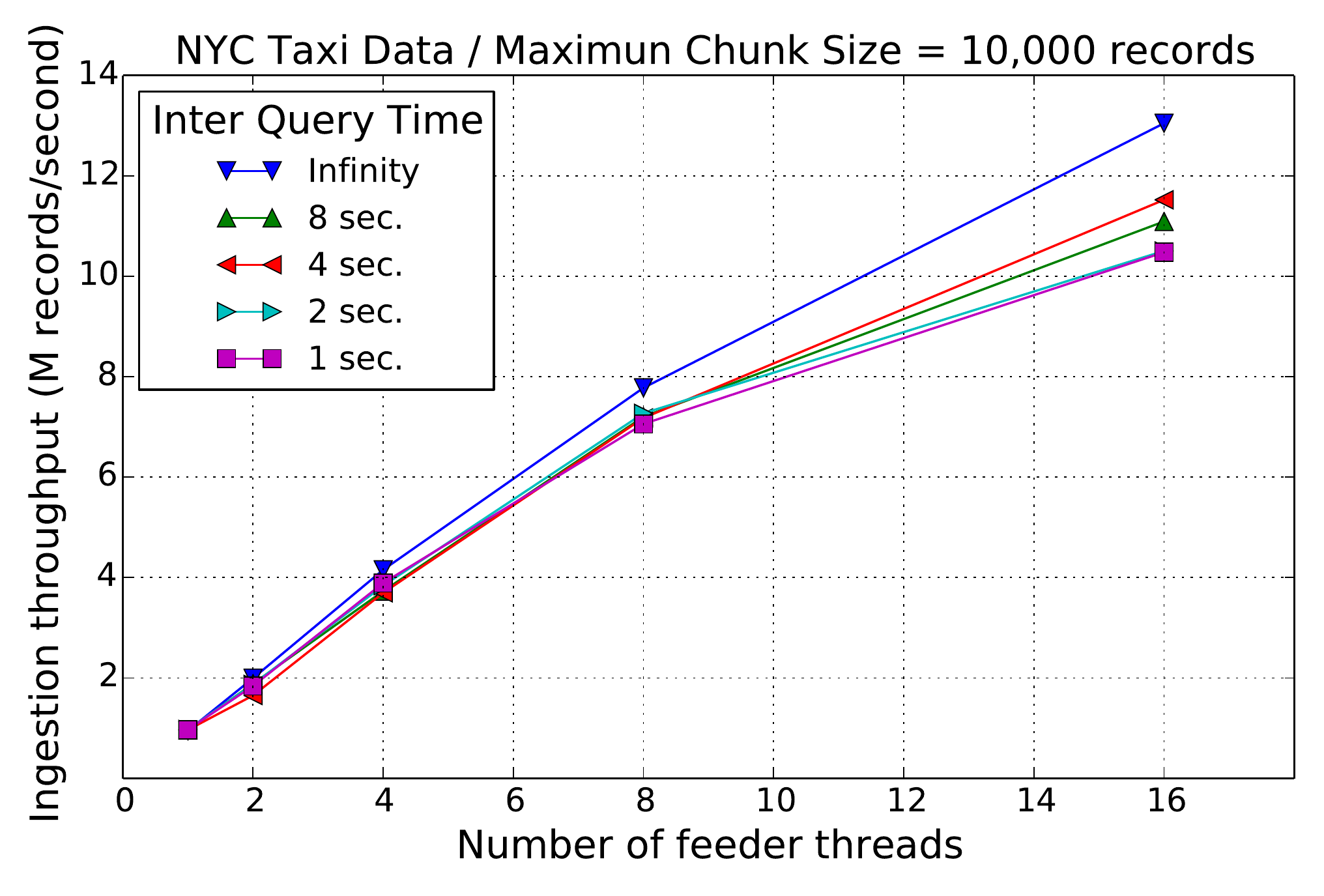}
\vspace*{-0.1in}
\caption{
Ingestion thread scaling of \mdds and the influence of queries.
}
\label{fig:mdds-thread-scaling}
\end{figure}

Figure~\ref{fig:mdds-thread-scaling} presents our results. 
We observe that \mdds is able to sustain $\sim$13M records/s of NYC taxi data
with 16 feeder threads.  
Also, it can sustain over 16M records/s of GHCN-Daily data with the same thread
count (not shown in the figure). 
By contrast, Percona Server (with the different engines), SQLite, and Druid (via
the Tranquility loader application) exhibit poor thread scaling and sustain much
lower ingestion rates; they only report up to 35K, 30K, and 55K records/s,
respectively.
Note that these systems perform worse in this multi-threaded case than when bulk
loading data.
To sum up, \mdds can sustain ingestion rates two orders of magnitude higher than
the competing systems -- at least 230$\times$ in the multi-thread scenario and
160$\times$ overall.

The previous result is obtained considering binary data. 
When ingesting CSV data with 16 feeder threads, \mdds only reaches over
2.7M~records/s; \ie 27$\times$ overall improvement.





%



In addition, Figure~\ref{fig:mdds-thread-scaling} shows two aspects of \mdds's ingestion
performance:
\begin{inparaenum}[(1)]
\item
\emph{how it varies with the number of feeder threads} (thread scaling), and 
\item
\emph{how queries influence it}.
\end{inparaenum}
The top line with downward triangle marker indicates the ingestion throughput
with no query processing (inter-query time $= \infty$). 
The other lines indicate the ingestion performance while processing queries in
sequence with different wait times between queries. 
The queries are selected uniformly at random from our test set for NYC taxi
data.
We report mean throughput over 5 runs.

Despite its superior performance, Figure~\ref{fig:mdds-thread-scaling} reveals
some limitations in our prototype.
When no queries are being processed \mdds's ingestion scales reasonably well up
to 16 threads, but not linearly.
We have observed that its ingestion throughput starts to plateau
after 20 threads on our test platform with hyperthreading enabled (not shown in
the figure).
In addition, queries can cause noticeable reduction of ingestion performance
(\eg $\sim$20\% with 16 feeder threads).
These limitations mainly result from contention on the R*-tree's lock and memory
(de)allocation, and addressing them are left as future work.

\subsubsection{Query Performance} \label{sec:query-perf}

Next we compare the query performance of \mdds and the other
systems.
The test queries and data schemas are listed in Appendix~\ref{sec:test-queries}. 
We report the average response time over 5 runs, after a warm-up phase of 2 runs
to avoid performance variability due to cold cache.

\begin{figure*}[tp]
\centering
\includegraphics[width=1.00\textwidth]{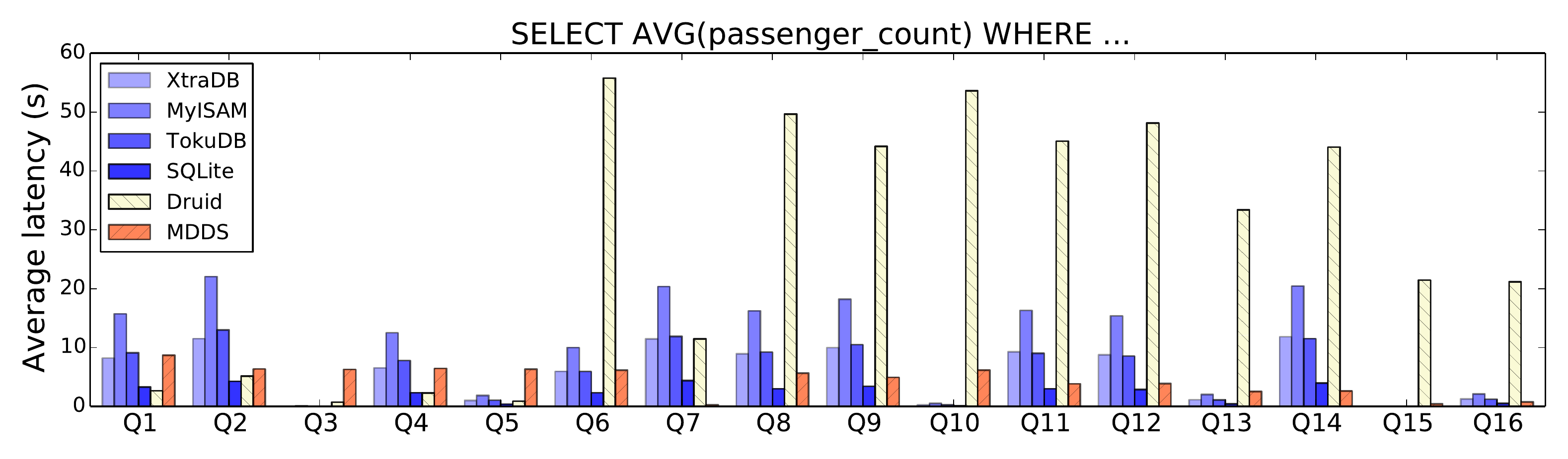}
\vspace*{-0.3in}
\caption{
Query response times for NYC taxi data. 
}
\label{fig:nyctaxi-query-resp-time}
\end{figure*}



Figure~\ref{fig:nyctaxi-query-resp-time} shows the results for NYC taxi data. 
SQLite surprisingly beats Percona Server's engines across all the queries.
But it is even more surprising that \mdds performs better than or comparably to
Percona Server's engines for 12 out of 16 queries on NYC taxi data. 
\mdds only does worse than Percona Server on queries Q3, Q5, Q10 and Q15, and
even outperforms SQLite for Q7 and Q14.
This result is unexpected because \mdds's query processor has not been heavily
optimized. 
We attribute the favorable query performance of SQLite and \mdds, which uses
SQLite's virtual table mechanism, to a nice match between 
(1)~their simple design and implementation, and
(2)~the simplicity of this test case involving only range queries served using a
single (multi-column or multi-dimensional) indexing structure.
Our results suggest SQLite is a reasonable choice to be the basis of \mdds's
query processor. 
Having being optimized over the years for read-heavy workloads, SQLite
expectedly reports better query performance than \mdds.
A plausible reason is that SQLite's cache is more effective than
\mdds's -- \mdds only caches data segments and not individual records.

We also observe in Figure~\ref{fig:nyctaxi-query-resp-time} that Druid is better
than \mdds and competitive with SQLite for two-dimensional queries (Q1--Q5), but
much worse than \mdds and SQLite for queries Q6--Q16, with 3- to 5-dimensional
ranges and one being on \texttt{pickup\_datetime}.
Druid exhibits long response times ($>$20s) for queries Q6 and Q8--Q16. 
This result suggests that the number of dimensional ranges may severely impact
Druid's query performance. 
Another possible explanation for such poor behavior is that
\texttt{pickup\_datetime} values are not stored in the timestamp column, but as
a dimension.

As expected, selectivity of matching data segments influences \mdds's query
performance.
\mdds tends to perform worse when queries are not selective in terms of
segments; \eg to serve Q3, Q5 and Q10 \mdds inspects all the segments (16,976).
By contrast, \mdds answers Q15 much faster, which involves only 
$\sim$10\% of the segments. 
Selectivity of matching records, on the other hand, appears to be less
determinant for \mdds's query performance; \eg \mdds takes about the same time
and inspects similar number of data segments to serve Q6 and Q10, but with very
different matching record counts
(Q6: 16,328 segments and $\sim$3.2M records; 
Q10: 16,976 segments and $\sim$75K records).


\begin{figure}[tp]
\centering
\includegraphics[width=1.05\columnwidth]{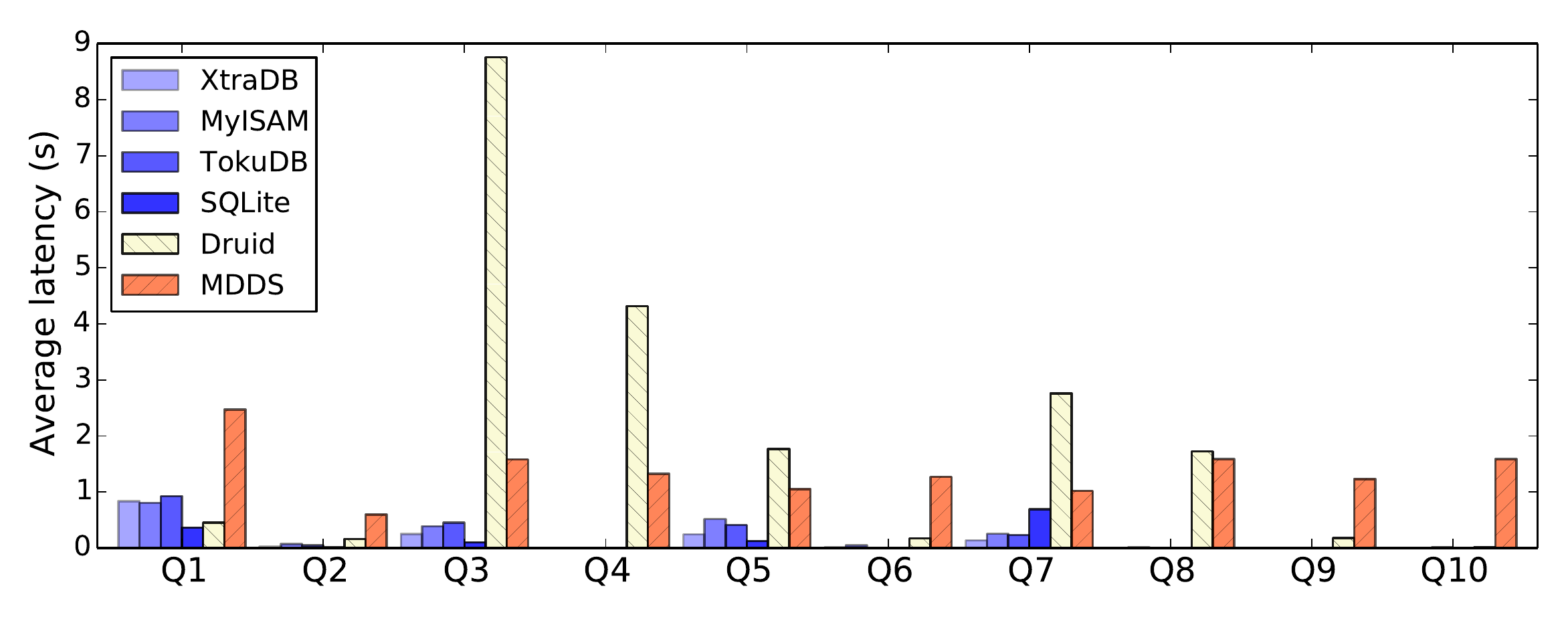}
\vspace*{-0.3in}
\caption{
Query response times for GHCN-Daily data. 
}
\label{fig:weather-query-resp-time}
\end{figure}

Figure~\ref{fig:weather-query-resp-time} reports the results for the set of more
realistic queries on GHCN-Daily data.
This test case is more favorable to Percona Server and SQLite because they use
multiple indices tailored to the queries; by contrast, \mdds uses a single
multi-dimensional index. 
Hence, as expected, these RDBMSs outperforms \mdds across all queries.
Note that these queries are not selective in terms of data segments -- \mdds
inspects 34\% of the segments to serve Q2, but 100\% (10K segments) or close for
Q1, Q3, and Q8, and between 60\% and 85\% for the rest.
As with NYC taxi data, SQLite is the best for most queries, albeit not
discernible in Figure~\ref{fig:weather-query-resp-time}.

In addition, \mdds outperforms Druid in half of the GHCN-Daily queries (Q3, Q4,
Q5, Q7 and Q8).
On the queries unfavorable to \mdds (\ie Q1, Q2, Q6, Q9 and Q10), 
Druid may benefit from the fact that no query includes a range on the
\texttt{time} field, not stored in the timestamp column but as a dimension. 
Moreover, Druid tends to perform worse than \mdds on queries
with 3 or more dimensional ranges.



%


\subsubsection{Storage Footprint} \label{sec:footprint}

\begin{table}[!tb]
\scriptsize
\centering
\begin{tabular}{p{0.38in}p{0.23in}p{0.3in}p{1.2in}}
\textbf{Storage Engine} & \textbf{NYC Taxi} & \textbf{GHCN Daily} & \multirow{2}{*}{\textbf{Notes}} \\
\hline
XtraDB  & 36.8  & 18.8  & Row format: COMPACT  \\ 
MyISAM  & 32.6  & 11.6  & Row format: Dynamic  \\ 
TokuDB  & 28.3  & 11.8  & Row format: tokudb\_zlib \\ 
SQLite  & 30.0  & 16.2  & ---\\
\multirow{2}{*}{Druid} & \multirow{2}{*}{11.9} & \multirow{2}{*}{7.2} & LZF, dictionary encoding, run-length encoding. \\
\mdds   & 22.8 & 4.9 & ---\\
\hline
\end{tabular}
\vspace*{-0.05in}
\caption{
Storage footprint (in GB).
}
\label{table:storage-footprint}
\end{table}

Table~\ref{table:storage-footprint} reports the space the systems use to store
both datasets, including indexing structures. 
\mdds stores NYC taxi data in 20\%-38\% less space than the RDBMSs, and only
about twice the space Druid occupies with heavy data compression (dictionary
encoding and LZF on data, and run-length encoding on bitmap lookup indices).
For GHCN-Daily data, \mdds occupies 26\%-42\% of the space required by the
RDBMSs, but remember they use four multi-column indices that increase their
storage needs.
Moreover, \mdds occupies 68\% of the space needed by Druid to store the GHCN-Daily
dataset. 
Thus, our results indicate that the superior query performance of the competing
systems over \mdds comes at the expense of larger storage footprint and much
lower ingestion rates, as shown in \S\ref{sec:1th-ingest-perf} and
\S\ref{sec:xth-ingest-perf}.





\subsection{Summary of Results} \label{sec:eval-summary}


Our comparative system evaluation (\S\ref{sec:sys-eval}) shows that \mdds can
ingest between 13M and 16M records/s with 16 feeder threads. 
Such ingestion rates are much higher than those of the competing systems -- at
least 160$\times$ higher considering binary data, and about 27$\times$ with CSV
data. 

With an unoptimized data retrieval path, \mdds reports query response times
comparable to or better than those of Percona Server, when using a single
multi-column index, and Druid.  
But, it is outperformed by SQLite in most queries on NYC Taxi Trip data for
which the latter uses a single multi-column index. 
Not surprisingly, \mdds is outperformed as well by the competing RDBMSs (Percona
Server and SQLite) when they use indices tailored for the test queries.  
Hence, for \mdds to offer competitive query performance its retrieval path
should be enhanced and optimized further.
These improvements should also reduce the degradation queries currently cause on
ingestion performance, especially at peak ingestion rates. 
Moreover, \mdds, using no compression technique, compares favorably in terms of
storage footprint against the other systems.

Our results also indicate that smaller data chunks tend to result in higher
ingestion throughput, and the number of dimensions, at least up to nine, does
not have a strong influence on the ingest rates.


In addition, our characterization of the segmentation schemes
(\S\ref{sec:dseg-eval}) shows that the kd-tree partitioning based scheme
(\S\ref{sec:kdtree-segmentation}) brings no clear advantage over the random
assignment scheme (\S\ref{sec:random-segmentation}). 
The random scheme is much simpler, 
offers ingestions rates that are higher and less susceptible to the number of
indexing dimensions, 
and 
results in comparable query performance. 
Note that compared to the random scheme, the kd-tree partitioning based scheme
produces less overlap between data segments at the expense of lower ingestion
performance.
But in our experiments such reduced overlap did not translate into noticeable
better query performance, contrary to what we initially hypothesized.

A key observation is that the packed kd-trees in data segments, along with the
kd-tree record iterator, offer important reductions in query response times. 
Hence, the time spent building the packed kd-trees and the storage space they
occupy are worth it.

\section{Related Work} \label{sec:relwork}

Numerous indexing structures for multidimensional data 
(\eg R-tree and its variants~\cite{rtreesbook2005}, 
kd-tree~\cite{bentley1975},
quadtree~\cite{finkel1974}, 
UB-tree~\cite{bayer1997}, 
PL-Tree~\cite{wang2013},
and PH-tree~\cite{zaschke2014}) 
have been proposed over the years, and some  
have been integrated into database systems to support application
domains like geomatics, data warehousing, and data mining. 

Traditional RDBMSs, such as Oracle~\cite{kothuri2002} and MySQL~\cite{mysql},
implement R-tree like indices for multidimensional data, typically targeting
geospatial applications. 
Thus, they often support few dimensions and are optimized for query processing
and not for fast data ingestion.

Researchers, inspired by NoSQL databases, have also proposed distributed
multidimensional data stores offering \emph{horizontal} scaling, fault
tolerance, and high ingestion rates (compared to RDBMSs).
These systems partition the data space into smaller subspaces (or shards) that
reside in slave nodes.
They often use a hierarchical two-level indexing structure, in which a
\emph{global index}, running across the master nodes, coarsely manages the
subspaces, and \emph{local indices}, running on each slave node, manage the data
records inside the subspaces.
Hence, subspaces along with the local indices resemble our data segments
(\S\ref{sec:dseg}).
The EMINC indexing structure~\cite{zhang2009}, albeit applied in a distributed
setting, is very similar to ours; it uses an R-tree as global index and kd-trees
as local indices.
The RT-CAN indexing scheme~\cite{wang2010} follows the same system architecture,
but builds the global index and local indices using \emph{content addressable
networks}~\cite{ratnasamy2001}.

Unlike EMINC and RT-CAN, the UQE-Index framework~\cite{ma2012} implements a
three-level indexing structure atop HBase key-value store~\cite{hbase}.
It is catered to IoT spatio-temporal data, carrying timestamps and locations.
The first indexing level handles time intervals using a B+-tree, the second
level indexes geographical areas with an R-tree, and  the third level indexes
data records inside each area using either an R-tree or a grid.
The first two indexing levels are global and coarse-grained, while the
third-level index is fine-grained and runs locally on the slave nodes.
Moreover, framework dedicates a number of nodes to the first two indexing levels
handling time intervals and geographical areas.

Other distributed multidimensional data stores, such as
MD-HBase~\cite{mdhbase2011} and Pyro~\cite{pyro2015}, are also build atop HBase
and implement an index layer that casts multidimensional data into a
unidimensional index and allows efficient query processing.
Pyro is tailored for spatio-temporal data and range queries, whereas MD-HBase
can handle higher dimensional data and k nearest neighbors (kNN) queries. 
By comparison, \mdds can index data using more than three dimensions, but does
not offer efficient processing of kNN queries; this is left as future work.

Finally, traditional online analytical processing (OLAP) systems mainly aim at
very quick responses to multidimensional analytical queries (\eg drill-downs and
aggregates) on large data volumes. 
Now they are also being designed to scale horizontally on commodity hardware and
ingest data at high velocity
(\eg~\cite{druid14,braun2015,ramnarayan2016,dehne2016,pedreira2016}). 
Druid~\cite{druid14}, introduced before, focuses on transactional events (logs),
in which timestamps is a key dimension and the data tends to be append-heavy. 
VOLAP~\cite{dehne2016}, proposed recently, is designed to
support up-to-date querying of high velocity data. 
It operates only in-memory, and supports data ingestion but not deletion. 
Like distributed data stores mentioned above, VOLAP partitions data into shards
that it stores in worker nodes. 
It also adopts a two-level indexing structure, in which the global index is
a modified PDC tree~\cite{dehne2012} and the local indices (for the shards in
the worker nodes) are Hilbert PDC trees~\cite{robillard2016}.
Moreover, VOLAP supports dimension hierarchies and offers a synchronization
scheme with configurable freshness. 
Cubrik~\cite{pedreira2016} is a distributed in-memory multidimensional DBMS,
developed at Facebook, capable of executing indexed OLAP operations. 
It partitions the data in ranges by every single dimension and stores the data
within containers, called \emph{bricks}, in an unordered and sparse fashion,
providing high ingestion rates and indexed access through any combination
of dimensions. 
The authors report an ingestion rate of $\sim$1M rows per second per node on a
cluster with only $\sim$20\% of CPU utilization. 
Cubrik's single-node ingestion performance is superior to that of Druid and
VOLAP, but much lower than the ingestion rates reached by \mdds.


%

\section{Conclusions} \label{sec:conc}

In this paper, we showed that it is possible to build a single-node
multidimensional data store able to ingest high-velocity sensor data while
offering reasonably good query performance.
The evaluation of our system prototype, \mdds, indicates that the adopted design
streamlines data ingestion and offers ingress rates at least 160$\times$ higher
than those of Percona Server, SQLite3, and Druid. 
This result suggests that there is potential for significant reductions in the
number of cluster nodes required to ingest high-velocity multidimensional data,
leading to more economical deployments. 
\mdds, with an unoptimized data retrieval path, also offers query response times
comparable to or better than those of Percona Server, using a single
multi-column index, and Druid on NYC Taxi Trip data. 
Furthermore, \mdds's storing footprint, without using any compression technique,
compares favorably against that of the other systems.

We also evaluated a kd-tree partitioning based scheme for segmenting incoming
streamed data. 
We verified that, compared to a random assignment scheme, the kd-tree
partitioning scheme produces less overlap between data segments at the expense
of lower ingestion performance. 
We, however, could not confirm the hypothesis that such reduced overlap yields
better query performance.
As a result, the random scheme is our first choice between both methods; it is
simpler, offers faster ingestion rates with comparable query performance on our
test queries, and its ingestion performance is not influenced by the number of
indexing fields (\ie dimensions).
Moreover, we observed that the packed kd-trees in data segments, along with the
kd-tree iterator, bring significant improvements on query performance.

All in all, we believe this paper provides useful information to practitioners
and researchers tasked with building a database system for high-velocity
multidimensional data.

As part of the future work, we plan to integrate \mdds into a distributed data
stream processing system like that depicted in Figure~\ref{fig:dist-sys-design}.
We want to understand the challenges of effectively coupling \mdds with a
single-node, high-throughput stream processing engine, like
StreamBox~\cite{streambox2017}, and creating a scalable system with these
building blocks. 
We also intend to overcome \mdds's current limitations, such as thread
scalability and ingestion performance degradation caused by queries, as well as
those listed in \S\ref{sec:mdds-arch}.
Finally, we plan to support more complex queries and continue investigating
swift segmentation schemes for ingress streamed data that can improve query
performance.


\setlength{\bibspacing}{\baselineskip}

\bibliographystyle{abbrv} 
\bibliography{main}

\begin{appendix}

\subsection{Schemas and Test Queries} \label{sec:test-queries}

Here we present the schemas and queries, in SQL, used in our comparative
evaluation (\S\ref{sec:sys-eval}).
For Druid, the schemas and queries were re-written in the proper JSON format.
\newline

\subsubsection{Schema for NYC Taxi Trip Data} \label{sec:nyctaxi-schema}
{
\scriptsize 
\begin{verbatim}
CREATE TABLE nyc_taxi_trip (
  medallion VARCHAR(32), hack_license VARCHAR(32),
  vendor_id VARCHAR(3), rate_code BIGINT,
  store_and_fwd_flag VARCHAR(1), 
  pickup_datetime BIGINT UNSIGNED, 
  dropoff_datetime BIGINT UNSIGNED, 
  passenger_count BIGINT UNSIGNED,
  trip_time_in_secs BIGINT SIGNED, trip_distance FLOAT,
  pickup_longitude FLOAT, pickup_latitude FLOAT,
  dropoff_long itude FLOAT, dropoff_latitude FLOAT)
ENGINE = [ InnoDB | MyISAM | TokuDB ];

CREATE INDEX idx1 ON nyc_taxi_trip (
  pickup_latitude, pickup_longitude, pickup_datetime,
  passenger_count, trip_time_in_secs);

\end{verbatim}
}

\subsubsection{Queries for NYC Taxi Trip Data} \label{sec:nyctaxi-queries}
{
\scriptsize 
\begin{verbatim}
Q1.  SELECT avg(passenger_count) FROM mdds_table WHERE 
     pickup_latitude >= 40.7645342324 AND 
     pickup_latitude <= 40.7735173889 AND 
     pickup_longitude >= -73.888983772 AND 
     pickup_longitude <= -73.8799803018;

Q2.  SELECT avg(passenger_count) FROM mdds_table WHERE 
     pickup_latitude >= 40.7567579105 AND 
     pickup_latitude <= 40.7657410671 AND 
     pickup_longitude >= -74.0104771822 AND 
     pickup_longitude <= -74.0014687259;

Q3.  SELECT avg(passenger_count) FROM mdds_table WHERE 
     pickup_latitude >= 40.8350997626 AND 
     pickup_latitude <= 40.8440829192 AND 
     pickup_longitude >= -73.9408829776 AND 
     pickup_longitude <= -73.9318997698;

Q4.  SELECT avg(passenger_count) FROM mdds_table WHERE 
     pickup_latitude >= 40.7313968589 AND 
     pickup_latitude <= 40.7403800154 AND 
     pickup_longitude >= -73.9281950202 AND 
     pickup_longitude <= -73.9191664656;

Q5.  SELECT avg(passenger_count) FROM mdds_table WHERE
     pickup_latitude >= 40.6395406815 AND
     pickup_latitude <= 40.648523838 AND
     pickup_longitude >= -73.7786922958 AND
     pickup_longitude <= -73.7695406582;

Q6.  SELECT avg(passenger_count) FROM mdds_table WHERE
     pickup_latitude >= 40.7766603868 AND
     pickup_latitude <= 40.7856435434 AND
     pickup_longitude >= -73.9849723371 AND
     pickup_longitude <= -73.9759755461 AND
     pickup_datetime >= 1357318152 AND
     pickup_datetime <= 1387058055;

Q7.  SELECT avg(passenger_count) FROM mdds_table WHERE
     pickup_latitude >= 40.7518744163 AND
     pickup_latitude <= 40.7608575729 AND
     pickup_longitude >= -73.9920231278 AND
     pickup_longitude <= -73.9830112587 AND
     pickup_datetime >= 1387425722 AND
     pickup_datetime <= 1389354509;

Q8.  SELECT avg(passenger_count) FROM mdds_table WHERE
     pickup_latitude >= 40.7390082969 AND
     pickup_latitude <= 40.7479914535 AND
     pickup_longitude >= -73.8975113729 AND
     pickup_longitude <= -73.8884894693 AND
     pickup_datetime >= 1358101957 AND
     pickup_datetime <= 1384534322;

Q9.  SELECT avg(passenger_count) FROM mdds_table WHERE
     pickup_latitude >= 40.7441121105 AND
     pickup_latitude <= 40.7530952671 AND
     pickup_longitude >= -73.9241726808 AND
     pickup_longitude <= -73.9151549391 AND
     pickup_datetime >= 1357228624 AND
     pickup_datetime <= 1380690924 AND
     passenger_count <= 5; 

Q10. SELECT avg(passenger_count) FROM mdds_table WHERE
     pickup_latitude >= 40.8095885444 AND
     pickup_latitude <= 40.8185717009 AND
     pickup_longitude >= -73.945678875 AND
     pickup_longitude <= -73.9366935177 AND
     pickup_datetime >= 1358303391 AND
     pickup_datetime <= 1388758328 AND
     passenger_count <= 5; 

Q11. SELECT avg(passenger_count) FROM mdds_table WHERE
     pickup_latitude >= 40.7405100269 AND
     pickup_latitude <= 40.7494931834 AND
     pickup_longitude >= -73.8848861323 AND
     pickup_longitude <= -73.875865478 AND
     pickup_datetime >= 1360644783 AND
     pickup_datetime <= 1378268507 AND
     passenger_count <= 6 AND
     trip_time_in_secs <= 1800; 

Q12. SELECT avg(passenger_count) FROM mdds_table WHERE
     pickup_latitude >= 40.7360481905 AND
     pickup_latitude <= 40.7450313471 AND
     pickup_longitude >= -73.9318252224 AND
     pickup_longitude <= -73.9228007954 AND
     pickup_datetime >= 1359423704 AND
     pickup_datetime <= 1377980909 AND
     passenger_count <= 2 AND
     trip_time_in_secs <= 3600;

Q13. SELECT avg(passenger_count) FROM mdds_table WHERE
     pickup_latitude >= 40.7973076091 AND
     pickup_latitude <= 40.8062907656 AND
     pickup_longitude >= -73.9652836263 AND
     pickup_longitude <= -73.9562951477 AND
     pickup_datetime >= 1358294025 AND
     pickup_datetime <= 1369831329 AND
     passenger_count <= 1 AND
     trip_time_in_secs <= 1800;

Q14. SELECT avg(passenger_count) FROM mdds_table WHERE
     pickup_latitude >= 40.7502443406 AND
     pickup_latitude <= 40.7592274972 AND
     pickup_longitude >= -74.0404189891 AND
     pickup_longitude <= -74.0314059325 AND
     pickup_datetime >= 1366484530 AND
     pickup_datetime <= 1381908856 AND
     passenger_count <= 2 AND
     trip_time_in_secs <= 1800;

Q15. SELECT avg(passenger_count) FROM mdds_table WHERE
     pickup_latitude >= 40.8966859263 AND
     pickup_latitude <= 40.9056690829 AND
     pickup_longitude >= -73.7811546197 AND
     pickup_longitude <= -73.7721573685 AND
     pickup_datetime >= 1361340581 AND
     pickup_datetime <= 1365432055 AND
     passenger_count <= 2 AND
     trip_time_in_secs <= 3600;

Q16. SELECT avg(passenger_count) FROM mdds_table WHERE
     pickup_latitude >= 40.64089501 AND
     pickup_latitude <= 40.6498781666 AND
     pickup_longitude >= -73.7831658104 AND
     pickup_longitude <= -73.7740165755 AND
     pickup_datetime >= 1385293736 AND
     pickup_datetime <= 1390390061 AND
     passenger_count <= 1 AND
     trip_time_in_secs <= 3600;

\end{verbatim}
}

\subsubsection{Schema for GHCN-Daily Data}
\label{sec:weather-schema}


{
\scriptsize 
\begin{verbatim}
CREATE TABLE IF ghnc_daily (
  time BIGINT UNSIGNED, station_id VARCHAR(16), 
  longitude FLOAT, latitude FLOAT, elevation FLOAT, 
  element_id INTEGER UNSIGNED, element_value INTEGER)
ENGINE = [ InnoDB | MyISAM | TokuDB ];


CREATE INDEX idx1 ON ghnc_daily(element_id, element_value); 
CREATE INDEX idx2 ON ghnc_daily(element_id, elevation); 
CREATE INDEX idx3 ON ghnc_daily(element_id, longitude, 
                                latitude, elevation); 
CREATE INDEX idx4 ON ghnc_daily(element_id, latitude); 

\end{verbatim}
}

\subsubsection{Queries for GHCN-Daily Data}
\label{sec:weather-queries}
{
\scriptsize 
\begin{verbatim}
Q1.  /* Average elevation where snow has occurred. */ 
     SELECT avg(elevation) FROM mdds_table WHERE 
     element_id = 1397641047;

Q2.  /* Mean lowest daily temperature where elevation is 
        above 4000 m. */ 
     SELECT avg(element_value)/10 FROM mdds_table WHERE 
     elevation > 4000 AND element_id = 1414351182;

Q3.  /* Average daily snowfall in Colorado, above 
        1000 m elevation. */
     SELECT avg(element_value) FROM mdds_table WHERE 
     longitude > -109.0 AND longitude < -102.0 AND 
     latitude > 37.0 AND latitude < 41.0 AND 
     elevation > 1000 AND element_id = 1397641047;

Q4.  /* Number of tornado events in Kansas. */
     SELECT count(*) FROM mdds_table WHERE 
     longitude > -102.0 AND longitude < -94.633 AND
     latitude > 37.0 AND latitude < 40.0 AND
     element_id = 1465135408;

Q5.  /* Mean daily average recorded temperature (tenths of a 
        degree) in the UK. */ 
     SELECT avg(element_value)/10.0 FROM mdds_table WHERE 
     longitude > -5.4327 AND longitude < 1.7688 AND
     latitude > 50.447 AND latitude < 54.6961 AND
     element_id = 1413568071;

Q6.  /* Average snow depth for Mount McKinley. */ 
     SELECT avg(element_value) FROM mdds_table WHERE 
     element_id = 1397643076 AND
     latitude >= 63.9 AND latitude <= 64.0 AND
     longitude > -152.282 AND longitude < -152.281;

Q7.  /* Maximum recorded temperature (degrees C) 
        in South Korea. */
     SELECT max(element_value)/10.0 FROM mdds_table WHERE
     element_id = 1414349144 AND
     longitude >= 126.227 AND longitude <= 129.6442 AND
     latitude >= 34.3117 AND latitude <= 38.1043;

Q8.  /* Number of observed temperatures above 
        54.4 degrees C. */
     SELECT count(*) FROM mdds_table WHERE 
     element_id = 1414349144 AND element_value > 544;

Q9. /* Number of smoke/haze events below sea-level. */
     SELECT count(*) FROM mdds_table WHERE 
     elevation < 0 AND elevation > -999.9 AND
     element_id = 1465135160;

Q10. /* Three highest elevations with dust, volcanic ash or 
        sand blowing. */
     SELECT distinct station_id, elevation FROM mdds_table WHERE
     element_id = 1465135159 ORDER BY elevation DESC LIMIT 3;

\end{verbatim}
}

\end{appendix}

\end{document}